\documentclass{ptephy_v1}

\preprintnumber{XXXX-XXXX} 

\usepackage{amsmath,amssymb}
\usepackage{subfig} 
\usepackage{url}
\usepackage{tikz}
\usepackage{float}
\usepackage{braket}
\usepackage{bm}

\begin{document}

\title{Axion cloud evaporation during inspiral of black hole binaries - the effects of backreaction and radiation -}


\author[1,*]{Takuya Takahashi}
\affil{Department of Physics, Kyoto University, Kyoto 606-8502, Japan \email{t.takahashi@tap.scphys.kyoto-u.ac.jp}}

\author[1]{Hidetoshi Omiya}

\author[1,2]{Takahiro Tanaka}
\affil{Center for Gravitational Physics, Yukawa Institute for Theoretical Physics, Kyoto University, Kyoto 606-8502, Japan}


\begin{abstract}%
Ultralight scalar fields such as axions can form clouds around rotating black holes (BHs) by the superradiant instability. 
It is important to consider the evolution of clouds associated with BH binaries for the detectability of the presence of clouds through gravitational wave signals and observations of the mass and spin parameters of BHs. 
The impact on the axion cloud due to the tidal perturbation from the companion in a binary system is first studied in Ref~\cite{Baumann:2019ztm}. 
Here, we re-examine this issue taking into account the following points. 
First, we study the influence of higher-multipole moments. Second, we consider the backreaction due to the angular momentum transfer between the cloud and the orbital motion. 
This angular momentum transfer further causes the backreaction to the hyperfine split 
through the change in geometry. 
Finally, we calculate the particle number flux to the infinity induced by the tidal interaction. 
As a result, we found that the scalar field is not reabsorbed by the BH. Instead, the scalar particles are radiated away to evaporate during the inspiral, irrespectively of the direction of the orbital motion, for almost equal mass binaries.
\end{abstract}

\maketitle

\section{Introduction}
Ultralight bosons such as axions are one of the best-motivated particles in the context of an extension of the Standard Model. The existence of plenty of axion-like particles is predicted by string theory~\cite{Arvanitaki:2009fg,Svrcek:2006yi}, and such particles are also a good candidate for the dark matter~\cite{Dine:1982ah,Preskill:1982cy,Abbott:1982af,Hui:2016ltb}. Interestingly, they can make an observable signature on cosmology and astrophysical phenomena. 
As one of them, we consider the observability of the existence of an ultralight scalar field associated with black holes (BHs).

A massive scalar field around a BH can be bound by gravitational interaction, and 
it can extract the rotation energy from a spinning BH through the superradiance~\cite{Brito:2020oca}. 
When the Compton wavelength of the scalar field is comparable to the radius of the BH, this energy extraction mechanism works most 
efficiently, and thus the field forms a macroscopic condensate around 
the BH. 
For astrophysical BHs, a scalar field with the mass in the range $10^{-20}\sim10^{-10}\mathrm{eV}$ has an appropriate Compton wavelength to form this condensate~\cite{Detweiler:1980uk,Dolan:2007mj}. 
In the following, we refer to such a scalar field simply as an axion, 
and its condensate formed around a BH as an {\it axion cloud}. 

One observational signature of the existence of an axion cloud 
is the appearance of a forbidden region in the parameter distribution of the 
mass and spin of BHs, owing to the spin-down due to the superradiance~\cite{Arvanitaki:2010sy,Brito:2014wla,Brito:2017zvb}. 
In addition, gravitational waves (GWs) emitted from an axion cloud and the modification of the gravitational waveform from a binary coalescence accompanied by an axion cloud are considered to be possible targets for observations~\cite{Arvanitaki:2014wva,Arvanitaki:2016qwi,Yoshino:2013ofa,Choudhary:2020pxy,Chung:2021roh,DeLuca:2021ite,Baumann:2018vus,Liu:2021llm}. 
From observations, we are now getting new constraints on the axion mass~\cite{Tsukada:2018mbp,Ng:2020ruv,LIGOScientific:2021jlr}.

The GW events observed so far are those originating from binary coalescences~\cite{Abbott:2016blz,TheLIGOScientific:2016qqj,LIGOScientific:2021djp}, 
and most of BHs whose spin is measured are belonging to a binary system. 
On the other hand, the tidal interaction from the companion in a binary 
greatly affects the evolution of the axion cloud during the inspiral phase. 
Axion clouds are initially dominated by the superradiant growing mode, 
but when the binary orbital frequency coincides with the difference in the phase velocities with another mode, a resonant level transition is induced~\cite{Baumann:2018vus}. 
The disruption of the cloud due to tidal interaction from the static perturber is also studied by numerical simulations~\cite{Cardoso:2020hca}.

If axions are transferred to a mode with a large decay rate, axions would be reabsorbed by the BH and the cloud is depleted before the merger~\cite{Berti:2019wnn,Takahashi:2021eso}. 
Furthermore, axions in the decaying mode typically have an angular momentum with the opposite sign to that of the central BH, and the reabsorption may accelerate the spin-down of the BH. As a result, the forbidden region in the mass and spin distribution of BHs might be modified. Therefore, to clarify the detectability of axions and to make the obtained constraints more robust, we need to understand the dynamics of axion clouds in binary systems more precisely.

The level transition of an axion cloud due to the tidal interaction is formulated in Ref.~\cite{Baumann:2019ztm}. 
We develop this work further in the following respects. 
In the previous works, the main focus was only on the effect of the leading quadrupole moment of the tidal potential. 
However, since the resonance caused by the effect of higher-multipole moments occurs earlier during the binary inspiral, 
it may dominate the evolution of clouds.
We also consider the backreaction due to the angular momentum transfer between the cloud and the orbital motion. 
We describe the transition between different axion energy eigenstates 
taking into account the backreaction to the orbital evolution using a non-linear model Hamiltonian, and derive the persistent rate of axions in a strongly non-linear regime. 
Furthermore, we propose a model to take into account the modification to 
the hyperfine split due to the backreaction of the angular momentum transfer,
and show this modification can drastically 
change the final fate of the transition of axion clouds. 
In addition, we examine a new channel that turns out to deplete the axion clouds; axion emission induced by the tidal perturbation. 
Once axions are transferred to an upper level by the higher-multipoles of the tidal field, they can be easily further excited to an unbound state. 
Then, axions evaporate away to infinity.

In this paper, we analyze axion clouds in the non-relativistic regime, 
assuming that the binary's orbit is quasi-circular and 
the orbital angular mementum is 
aligned to the BH spin, for simplicity. 
Also, we neglect the axion self-interaction, 
which may play an important role in the evolution of axion clouds~\cite{Baryakhtar:2020gao, Omiya:2020vji}. 
Hence, we consider the simplest model of a massive scalar field 
and assume that there is nothing that prevents the growth of clouds 
until the superradiance condition is saturated by the spin-down 
of the central BH. 
We believe that this work will give a foundation of the analysis 
of more complicated cases.

This paper is organized as follows. In Sec.~\ref{Sec2}, 
we review how the level transition of axion clouds is described 
and show the importance of higher multipole moments of the tidal potential. In Sec.~\ref{Sec3}, we analyze the effect of 
the backreaction of the angular momentum transfer in detail. In Sec.~\ref{Sec4}, we calculate the particle number flux to the 
infinity. Finally, In Sec.~\ref{Sec5}, 
we summarize our results and give a brief discussion about their implications. 
In the rest of this paper, we use the unit with $G=c=\hbar=1$.

\section{Level transition of axion clouds}\label{Sec2}
In this section, we first review the level transition of axion clouds 
induced by the tidal perturbation due to the binary companion, 
formulated in Ref.~\cite{Baumann:2019ztm}. 
Then, we show that higher-multipole moments of the tidal perturbation actually play an important role in the evolution of clouds during the inspiral phase.

\subsection{Hamiltonian}
The equation of motion of a real scalar field (axion) of mass $\mu$ around a rotating BH is 
\begin{equation}\label{eomex}
\left(g^{\mu\nu}\nabla_{\mu}\nabla_{\nu}-\mu^2\right)\phi=0~,
\end{equation}
where $g_{\mu\nu}$ is the Kerr metric. Let the BH mass be $M$ and the angular momentum be $J=aM$. 
In the non-relativistic limit, it is appropriate to start with the ansatz
\begin{equation}
\phi=\frac{1}{\sqrt{2\mu}}\left(e^{-i\mu t}\psi+e^{i\mu t}\psi^{\ast}\right)~.
\end{equation}
and $\psi$ is a complex scalar function whose timescale of variation is much larger than $\mu^{-1}$. 
With this ansatz, the equation of motion (\ref{eomex}) is rewritten as 
\begin{equation}\label{eomNR}
i\frac{\partial}{\partial t}\psi=\left(-\frac{1}{2\mu}\nabla^2-\frac{\alpha}{r}\right)\psi~, 
\end{equation}
at the leading order in $r^{-1}$. 
Here, we have introduced the gravitational fine structure constant
\begin{equation}
\alpha=M\mu~\simeq0.2\left(\frac{M}{30M_{\odot}}\right)\left(\frac{\mu}{10^{-12}\mathrm{eV}}\right)~.
\end{equation}
Eq.\eqref{eomNR} is formally equivalent to the Schr$\mathrm{\ddot{o}}$dinger equation and the bound state solution is given by
\begin{equation}
\psi_{nlm}=e^{-i(\omega_{nlm}-\mu) t}R_{nl}(r)Y_{lm}(\theta,\varphi)~,
\end{equation}
where $R_{nl}(r)$ represents the radial wave function of the hydrogen atom (Appendix.~\ref{mathfunc}) and 
$Y_{lm}(\theta,\varphi)$ is the spherical harmonics. 
The radius corresponding to the Bohr radius for a hydrogen atom is $r_{0}=M/\alpha^2$, and this approximation is well justified for $\alpha\ll1$. In the non-relativistic approximation, 
the eigenfrequency is given by~\cite{Baumann:2019eav,Baumann:2018vus,Detweiler:1980uk}
\begin{equation}
\omega_{nlm}=(\omega_{R})_{nlm}+i(\omega_{I})_{nlm}~,
\end{equation}
with 
\begin{align}
(\omega_{R})_{nlm}&=\mu\left(1-\frac{\alpha^2}{2n^2}-\frac{\alpha^4}{8n^4}+\frac{(2l-3n+1)\alpha^4}{n^4(l+1/2)}+\frac{2m\alpha^5}{n^3 l(l+1/2)(l+1)}\frac{a}{M}\right)~, \label{Ene} \\
(\omega_{I})_{nlm}&=2(r_{+}/M)C_{nlm}(a,\alpha)(m\Omega_{H}-\omega)\alpha^{4l+5}~, \label{omeI}
\end{align}
where $r_{+}$ is the horizon radius, $\Omega_{H}=a/2Mr_{+}$ is the angular velocity of the BH horizon and explicit form of $C_{nlm}(a,\alpha)$ is given in Ref~\cite{Detweiler:1980uk}. 
Note that the labels of the eigenstates that we can take are restricted to $n\geq1, l\leq n-1,|m|\leq l$. 
As you can see from Eq.\eqref{omeI}, the mode satisfying $\omega_{R}<m\Omega_{H}$ is a growing mode. 
We can consider the situation in which axions occupy the fastest growing mode initially. 
In the following, we assume that there is no mechanism to terminate the growth of the cloud. Then, the growth of the cloud will saturate, when the the BH spin is reduced to satisfy 
the condition for the critical spin
\begin{equation}
\frac{a}{M}=\frac{a_{\mathrm{crit}}}{M}\equiv\frac{4m\alpha}{m^2+4\alpha^2}~. 
\end{equation}

When a BH with an axion cloud forms a binary system, the tidal 
perturbation from the binary companion of mass $M_{\ast}$ at $(R_{\ast},\Theta_{\ast},\Phi_{\ast})$ can be described by 
the additional potential~\cite{Baumann:2018vus} 
\begin{equation}\label{Vast}
V_{\ast}(t)=-M_{\ast}\mu\sum_{l_{\ast}=2}^{\infty}\sum_{|m_{\ast}|\leq l_{\ast}}\frac{4\pi}{2l_{\ast}+1}Y_{l_{\ast}m_{\ast}}^{\ast}(\Theta_{\ast},\Phi_{\ast})Y_{l_{\ast}m_{\ast}}(\theta,\varphi)\left(\frac{r^{l_{\ast}}}{R_{\ast}^{l_{\ast}+1}}\theta(R_{\ast}-r)+\frac{R_{\ast}^{l_{\ast}}}{r^{l_{\ast}+1}}\theta(r-R_{\ast})\right)
\end{equation}
in Eq.(\ref{eomNR}). 
The orbital angular velocity of the companion can be defined by $\dot{\Phi}_{\ast}(t)=\pm\Omega(t)$, 
and the upper (lower) sign represents the case of co-rotating (counter-rotating) orbits relatively to the central BH spin. 
For simplicity, we focus on the case in which the binary orbit is on the equatorial plane, $\Theta=\pi/2$. 

We denote the $i$-th eigenstate of an axion as $\ket{i}$, and a general bound state can be written by a superposition of eigenstates, {\it i.e.}, $\ket{\psi(t)}=\sum_{i}\tilde{c}_{i}(t)\ket{i}$. 
Since the tidal interaction acts as an periodically oscillating 
tiny external field and becomes important only when the orbital angular velocity coincides with the difference between the phase velocities 
of the two levels before and after transition.  
Therefore, it is sufficient to consider a two level system, neglecting all the other levels. 
$|\tilde{c}_{i}(t)|^2$ represents the normalized particle number occupying the level $i$, and its time evolution is given by
\begin{gather}
i\frac{\mathrm{d}\tilde{c}_{i}}{\mathrm{d}t}=\sum_{j=1,2}\tilde{\mathcal{H}}_{ij}\tilde{c}_{j}~, \\
\tilde{\mathcal{H}}=\left(\begin{array}{cc}
-\frac{\Delta E}{2} & \eta e^{i\Delta m\Phi_{\ast}(t)} \\
\eta e^{-i\Delta m\Phi_{\ast}(t)} & \frac{\Delta E}{2}
\end{array}\right)~, \label{Hori}
\end{gather}
where $\Delta E=(\omega_{R})_{2}-(\omega_{R})_{1}$ and $\Delta m=m_{2}-m_{1}$. 
Writing
\begin{equation}
V_{\ast}(t)=\sum_{l_{\ast}m_{\ast}}V_{\ast,\l_{\ast}m_{\ast}}e^{-im\Phi_{\ast}(t)}~,
\end{equation}
we parameterize the amplitude of perturbation by $\eta\equiv|\braket{1|V_{\ast,l_{\ast}m_{\ast}}|2}|$, whose explicit expression is 
\begin{align}
\frac{\eta}{\Omega}=&\left|\frac{q}{(1+q)^{(1+l_{\ast})/3}}\frac{(M\Omega)^{(2l_{\ast}-1)/3}}{\alpha^{2l_{\ast}-1}}\frac{4\pi}{2l_{\ast}+1}Y_{l_{\ast}m_{\ast}}\left(\Theta_{\ast},\Phi_{\ast}\right)\ I_{\tilde{r}}^{\mathrm{in}}\times I_{A}\right. \notag \\
&\left.+ q(1+q)^{l_{\ast}/3}\frac{\alpha^{2l_{\ast}+3}}{(M\Omega)^{(2l_{\ast}+3)/3}}\frac{4\pi}{2l_{\ast}+1}Y_{l_{\ast}m_{\ast}}\left(\Theta_{\ast},\Phi_{\ast}\right)\ I_{\tilde{r}}^{\mathrm{out}}\times I_{A}\right|~,  \label{eta}
\end{align}
with
\begin{gather}
I_{\tilde{r}}^{\mathrm{in}}=\int_{0}^{R_{\ast}/r_{0}}d\tilde{r}\ \tilde{r}^{2+l_{\ast}}\tilde{R}_{n_{1}l_{1}}\tilde{R}_{n_{2}l_{2}}~, \quad
I_{\tilde{r}}^{\mathrm{out}}=\int_{R_{\ast}/r_{0}}^{\infty}d\tilde{r}\ \tilde{r}^{1-l_{\ast}}\tilde{R}_{n_{1}l_{1}}\tilde{R}_{n_{2}l_{2}}~,\\
I_{A}=\int \sin\theta d\theta d\varphi\ Y_{l_{1}m_{1}}^{\ast}Y_{l_{\ast}m_{\ast}}Y_{l_{2}m_{2}}~,
\end{gather}
where $\tilde{r}=r/r_{0}$, $\tilde{R}_{nl}=r_{0}^{3/2}R_{nl}$. 
Here we defined the mass ratio by
\begin{equation}
q=\frac{M_{\ast}}{M}~.
\end{equation}
Performing the unitary transformation with the transformation matrix $
\mathcal{U} = \mathrm{diag} ( e^{i\Delta m\Phi_{\ast}(t)/2}$,
$ e^{-i\Delta m\Phi_{\ast}(t)/2} ) $, 
we can remove the rapidly oscillating terms in Eq.\eqref{Hori}. 
In the new frame, the coefficients and Hamiltonian are, respectively, transformed to $c(t)=\mathcal{U}^{-1}\tilde{c}(t)$ and $\mathcal{H}=\mathcal{U}^{\dagger}\mathcal{H}\mathcal{U}-i\mathcal{U}^{\dagger}\dot{\mathcal{U}}$. 
As a result, we can describe the level transition due to the tidal interaction by the following equation with the Hamiltonian;
\begin{gather}
i\frac{\mathrm{d}c_{i}}{\mathrm{d}t}=\sum_{j=1,2}\mathcal{H}_{ij}c_{j}~, \\
\mathcal{H}=\left(\begin{array}{cc}
\pm\frac{\Delta m}{2}(\Omega(t)-\Omega_{0}) & \eta \\
\eta & \mp\frac{\Delta m}{2}(\Omega(t)-\Omega_{0})
\end{array}\right)~, \label{H} 
\end{gather}
where we defined the resonance frequency by
\begin{equation}\label{resO}
\Omega_{0}=\pm\frac{\Delta E}{\Delta m}~.
\end{equation}

\subsection{Linear orbital evolution}
For the binary BH system, the orbital angular velocity gradually increases due to the GW radiation reaction. During the inspiral phase, the time evolution of the orbital angular velocity is given by~\cite{PhysRev.131.435,Blanchet:2013haa}
\begin{gather}
\frac{\mathrm{d}\Omega}{\mathrm{d}t}=\gamma\left(\frac{\Omega}{\Omega_{0}}\right)^{11/3}~, 
\end{gather}
with 
\begin{gather}
\frac{\gamma}{\Omega_{0}^2}=\frac{96}{5}\frac{q}{(1+q)^{1/3}}(M\Omega_{0})^{5/3}~.
\end{gather}
If we approximate $\Omega(t)$ linearly around the resonance frequency;
\begin{equation}
\Omega(t)=\Omega_{0}+\gamma t~,
\end{equation}
the Hamiltonian  \eqref{H} becomes
\begin{equation}
\mathcal{H}=\left(\begin{array}{cc}
\pm\frac{\Delta m}{2}\gamma t & \eta \\
\eta & \mp\frac{\Delta m}{2}\gamma t
\end{array}\right)~.
\label{Hamiltonian2}
\end{equation}
The problem defined by the the Hamiltonian \eqref{Hamiltonian2} 
is known as {\it the Landau Zener problem} \cite{10011873546,Zener:1932ws} and the dynamics of this system is characterized by the parameter
\begin{equation}\label{zdef}
z\equiv\frac{\eta^2}{|\Delta m|\gamma}~,
\end{equation}
which measures the adiabaticity of the transition. 
This set of equations have an exact solution, and we can calculate the persistent rate $\Gamma=|c_{1}(+\infty)|^2/c_{1}(-\infty)|^2$ analytically. 
With the initial condition $c_{1}(-\infty)=1$ and $c_{2}(-\infty)=0$, 
the persistent rate is given by~\cite{Baumann:2019ztm} 
\begin{equation}\label{lPR}
\Gamma=\exp(-2\pi z)~.
\end{equation}
This means that almost all axions are transferred to another mode in the adiabatic limit, $z\gg1$.

\subsection{Higher-multipole moments}\label{SecHM}
In Ref.~\cite{Baumann:2019ztm}, the main focus is on the effect of leading quadrupole tidal perturbation; $l_{\ast}=2$ in Eq.(\ref{Vast}). 
In this case, only the transition with $\Delta m\leq 2$ is allowed because of 
the selection rules reflecting the conservation of the angular momentum; $-m_{1}+m_{\ast}+m_{2}=0$ and $|m_{\ast}|\leq l_{\ast}$. 
If we take into account the effect of higher-multipole moments $l_{\ast}>2$, 
a wider class of transitions can occur. 
From Eq.(\ref{resO}), we can see that, if we have the same order of the difference 
between the energy levels, $\Delta E$, the larger $\Delta m$ and $l_{\ast}$, the smaller resonance frequency $\Omega_{0}$ becomes.  
Therefore, if the magnitude of the perturbation of higher multipole moments 
is sufficiently large, 
since their resonances occur earlier, they might dominate the transition. 
In the following, we examine to which mode the axions that initially occupy $\ket{n_{1}l_{1}m_{1}}=\ket{211}$ make the first transition effectively.

For the co-rotating case, the transition with $\Delta E/\Delta m>0$ occurs. 
In particular, the transition between two levels separated by the hyperfine split, {\it i.e.}, $\ket{211}\to\ket{21\mbox{-1}}$,
is possible. 
As one can see from Eq.(\ref{Ene}), this energy gap $|\Delta E|$ is very tiny, 
since it is suppressed by a factor of $\mathcal{O}(\alpha^4)$ compared to the gap 
between different $n$ modes, which is $\mathcal{O}(\alpha^2)$. 
For this reason, even if we take higher-multipole moments into account, 
binaries evolution hits the frequency for the transition due to the 
quadrupole moment first.

\begin{figure}[t]
 \begin{tabular}{cc}
 \begin{minipage}[t]{0.5\hsize}
	\centering
	\includegraphics[keepaspectratio,scale=0.6]{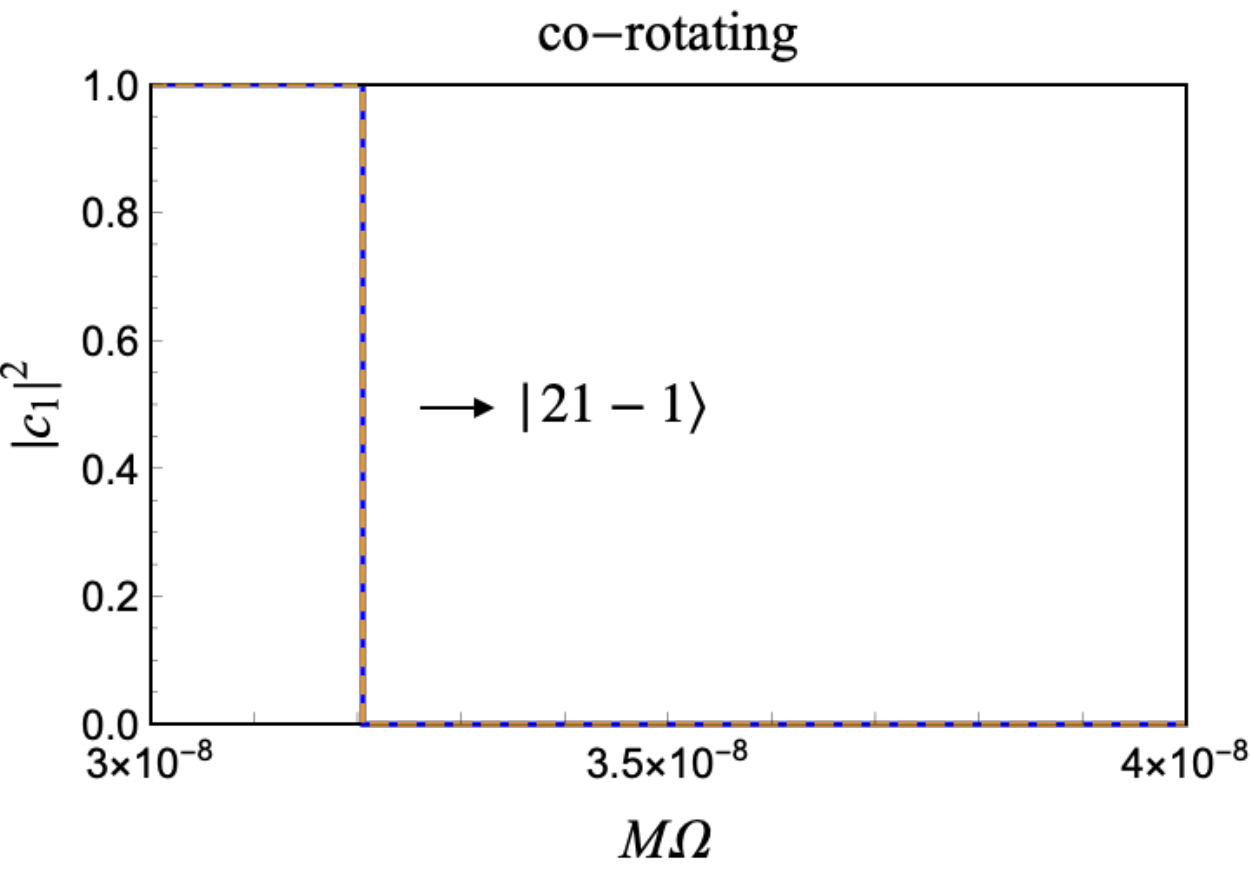}
\end{minipage} &
\begin{minipage}[t]{0.5\hsize}
        \centering
	\includegraphics[keepaspectratio,scale=0.6]{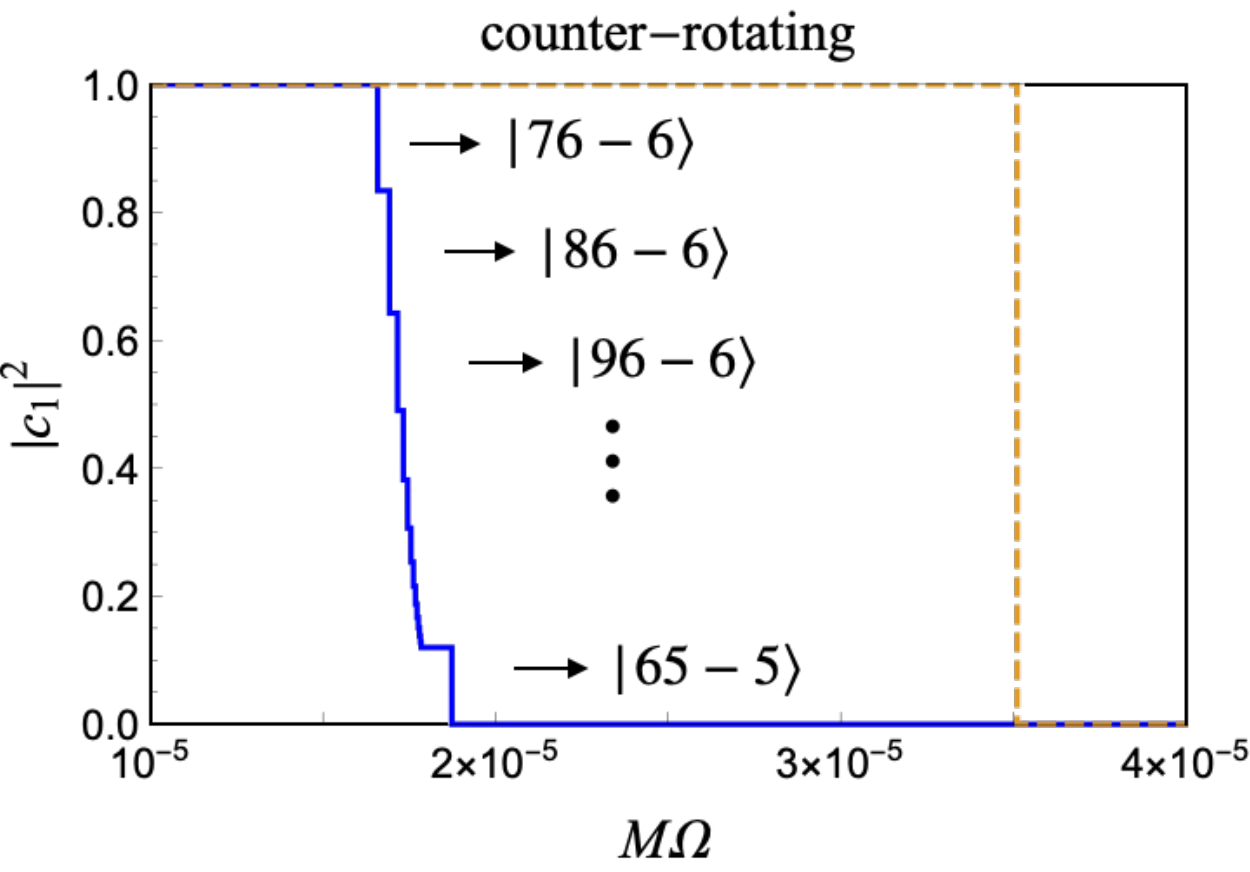}
\end{minipage}
\end{tabular}
	\caption{Evolution of the normalized particle number occupying $\ket{211}$, $|c_{1}|^2$,  for the co-rotating case (left) and the counter-rotating case (right) with $\alpha=0.1,q=1$. The blue solid curves show the transition taking into account the higher-multipole moments. The yellow dashed curves show the transition due to the quadrupole moment. For the co-rotating case, they are the same. While, for counter-rotating case, $(l_{\ast}=) 7$-th moment first gives a non-negligible contribution to the transition. The right side of the arrows shows the transition destinations for each step.}
	\label{figHMtr}
\end{figure}

For the counter-rotating case, the transition with $\Delta E/\Delta m<0$ occurs. 
In this case, only the transition to the level with a higher $n$ occurs. 
Since $\Delta E$ is the same order for all transitions 
that are associated with the change of $n$, 
the transition with a large $|\Delta m|$ works earlier. 
The question that we need to answer is which is the largest $l_{\ast}$ that 
starts to contribute to the transition effectively, 
because the amplitude of perturbation $\eta$ is suppressed like $\propto R_{\ast}^{-(l_{\ast}+1)}$. 
As an example, we show the evolution of the particle number 
$|c_{1}|^2$ for $\alpha=0.1, q=1$ in Fig.~\ref{figHMtr}. 
This evolution takes into account the higher-multipole moments, 
and $|c_1|^2$ is reduced by multiplying the persistent rate \eqref{lPR} 
at each resonance frequency for all transitions satisfying the selection rules.  
For this set of parameters, the $(l_{\ast}=)7$-th moment is the first one 
that gives a non-negligible contribution to the transition, 
and a part of axions in $\ket{211}$ are transferred to $\ket{76-6}$. 
Successively, the axions remaining in $\ket{211}$ are gradually transferred to the levels with a high $n$ ({\it e.g.}, $\ket{86-6},\ket{96-6},\cdots$). 

\begin{figure}[t]
 \begin{tabular}{cc}
 \begin{minipage}[t]{0.5\hsize}
	\centering
	\includegraphics[keepaspectratio,scale=0.5]{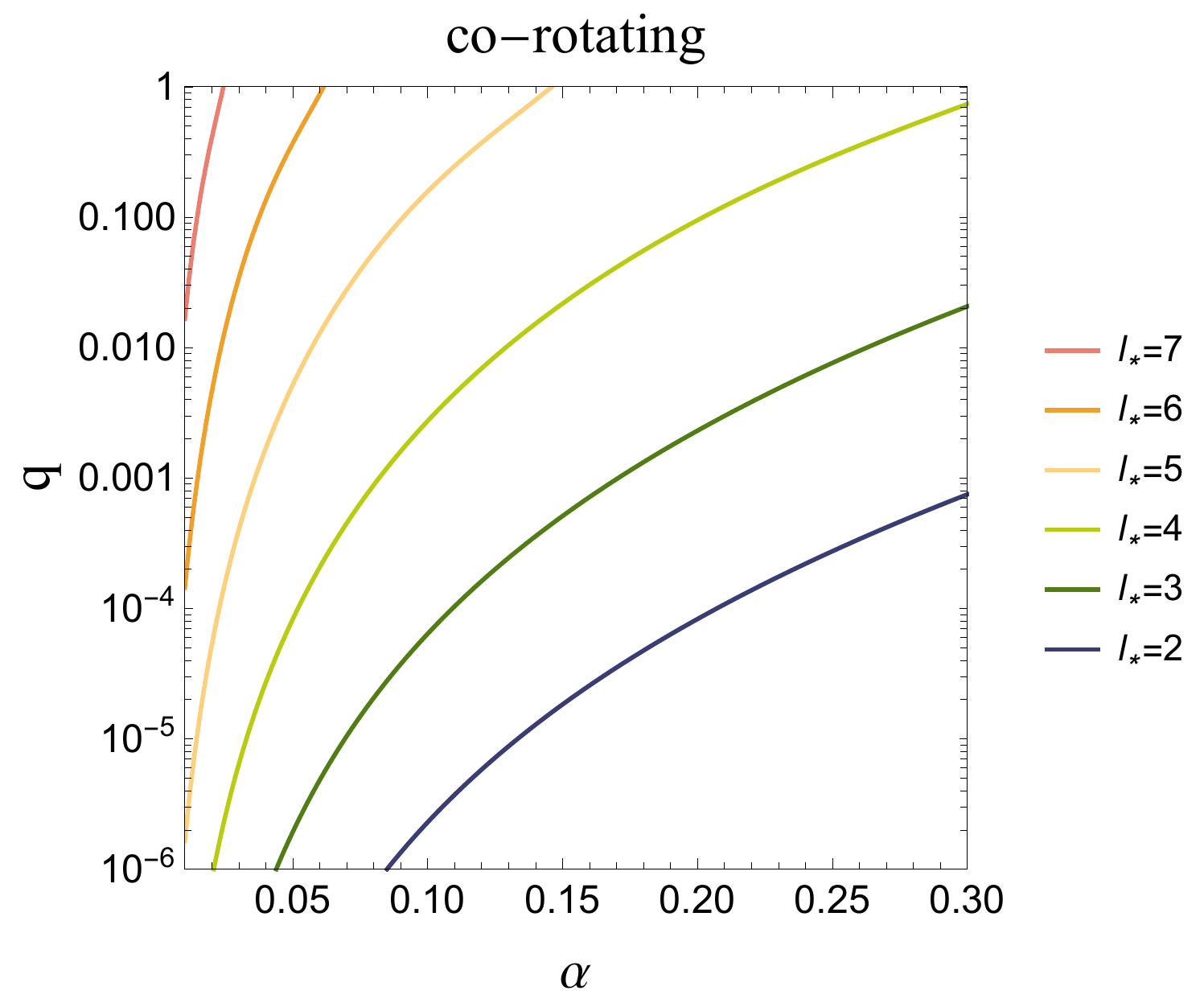}
\end{minipage} &
\begin{minipage}[t]{0.5\hsize}
        \centering
	\includegraphics[keepaspectratio,scale=0.5]{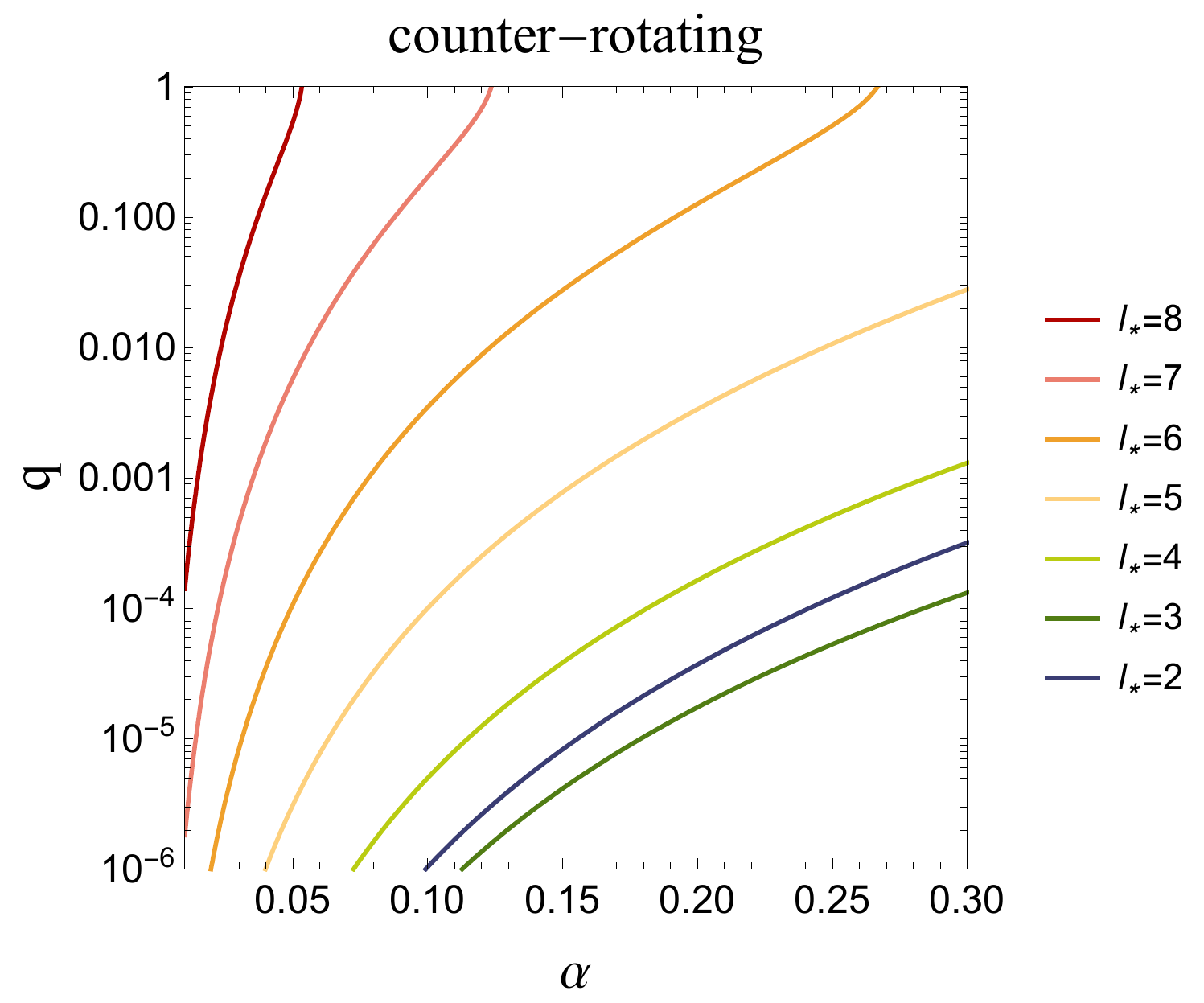}
\end{minipage}
\end{tabular}
	\caption{Boundary curves above which the $l_{\ast}$-th moment can give a non-negligible contribution to the transitions for the co-rotating case (left) and for the counter-rotating case (right). The solid curves show the boundaries where the adiabaticity $z$ is equal to 0.01 for the transition with the smallest resonance frequency for the respective specified values of $l_{\ast}$. 
	In the parameter region above the curve, the transition due to the $l_{\ast}$-th moment effectively works.}
	\label{figWhichl}
\end{figure}

Next, we discuss the dependence on the parameter set $(\alpha,q)$. 
For a given $l_{\ast}$ and an initial state $(n_1,l_1,m_1)$, 
we consider the combinations of $(n_{2},l_{2},m_{2})$ for the transition destination that minimizes the resonance frequency, {\it i.e.} the energy difference $\Delta E$ 
divided by $\Delta m$. 
First, to maximize $\Delta m$, $m_{\ast}$ is determined to be $\pm l_{\ast}$, then we have $m_{2}=m_{1}\pm l_{\ast}$. 
From Eq.(\ref{Ene}), $\omega_{R}$ is minimized for the minimum $n$ and $l$ that we can take. 
Hence, $\Delta E$ is minimized for $n_{2}=|m_{2}|+1$ and $l_{2}=|m_{2}|$, 
assuming that the hyperfine split is negligible. 
Considering the transition from $\ket{211}$ and calculating the persistent rate from the adiabaticity $z$, 
we determine the region in which the $l_{\ast}$-th moment can contribute to 
the transition in the parameter space of $(\alpha,q)$ as shown in Fig.~\ref{figWhichl}
\footnote{For the counter-rotating case, we consider the transition $\ket{211}\to\ket{31\mbox{-1}}$ with $l_{\ast}=2$ moment.}. 
We set the threshold as $z=0.01$. In the parameter region below the curves, the transition rate, which is determined by $z$, is so small that one would be able to neglect its contribution to the cloud depletion. 
As one can see from this figure, it is important to take into account the higher-multipole moments, at least for almost equal mass binaries.

Here, we confirmed that higher-multipole moments play an important role in the level transition during the binary inspiral. However, as we will see in the next section, we will find that the backreaction due to the angular momentum transfer cannot be ignored for both co-rotating and counter-rotating cases. Therefore, we should understand that the discussion presented here is not the final conclusion about the transition destination. 

\section{Backreaction}\label{Sec3}
In this section, we investigate the backreaction caused by the angular momentum transfer associated with the level transition. When higher-multipole moments are considered, the transitions occur at smaller resonance frequencies compared with the case in which only the quadrupole moment is considered. Therefore, since the change rate of the orbital angular momentum is smaller, the effect of the backreaction becomes more important. We also investigate the effect of the backreaction on the hyperfine split due to the angular momentum transfer between the axion cloud and the orbital motion. In addition, we analytically derive a formula for the persistent rate taking into account these backreaction effects.

\subsection{Non-linear Hamiltonian}
\begin{figure}[t]
 \begin{tabular}{cc}
 \begin{minipage}[t]{0.5\hsize}
	\centering
	\includegraphics[keepaspectratio,scale=0.53]{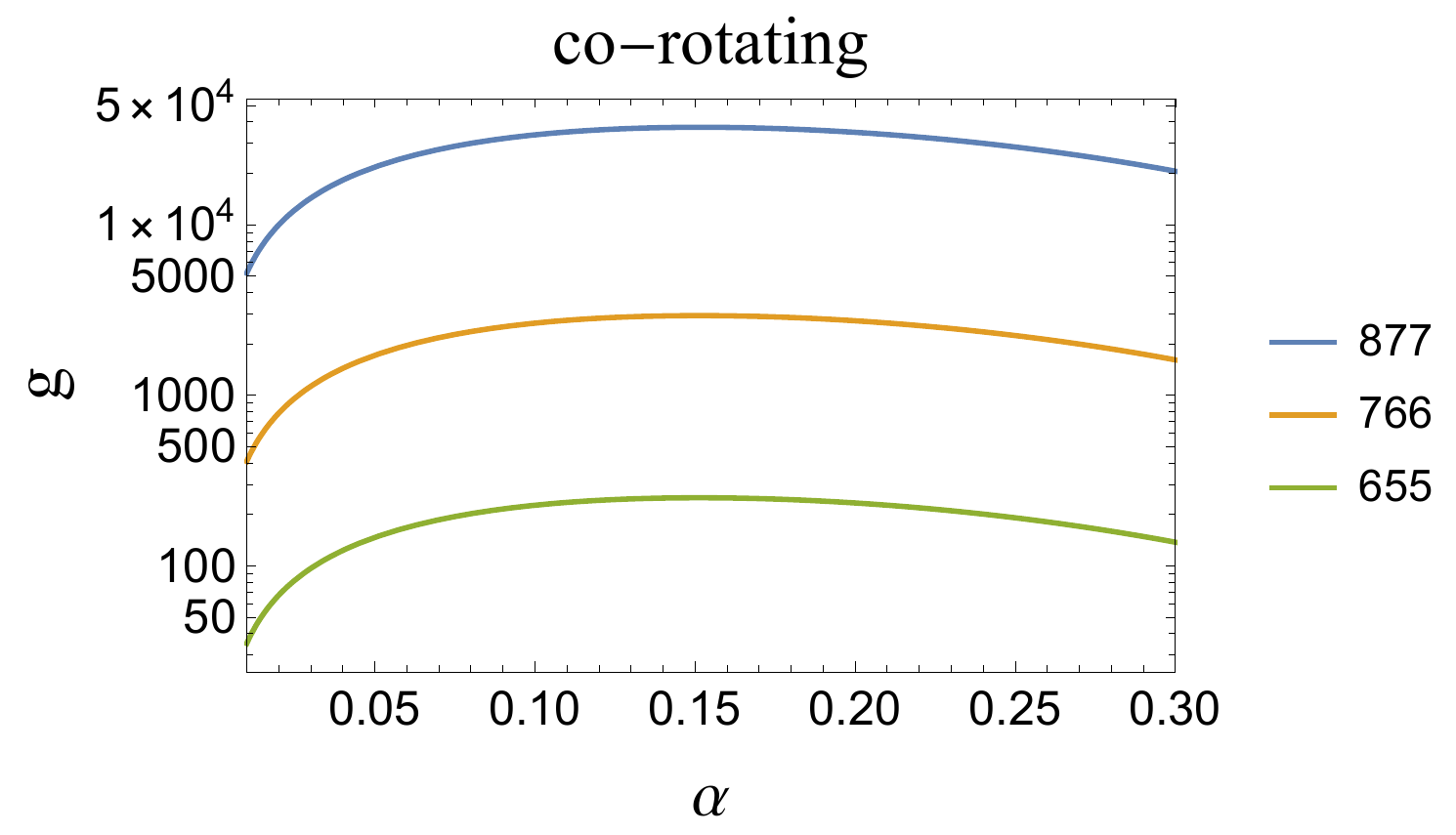}
\end{minipage} &
\begin{minipage}[t]{0.5\hsize}
        \centering
	\includegraphics[keepaspectratio,scale=0.53]{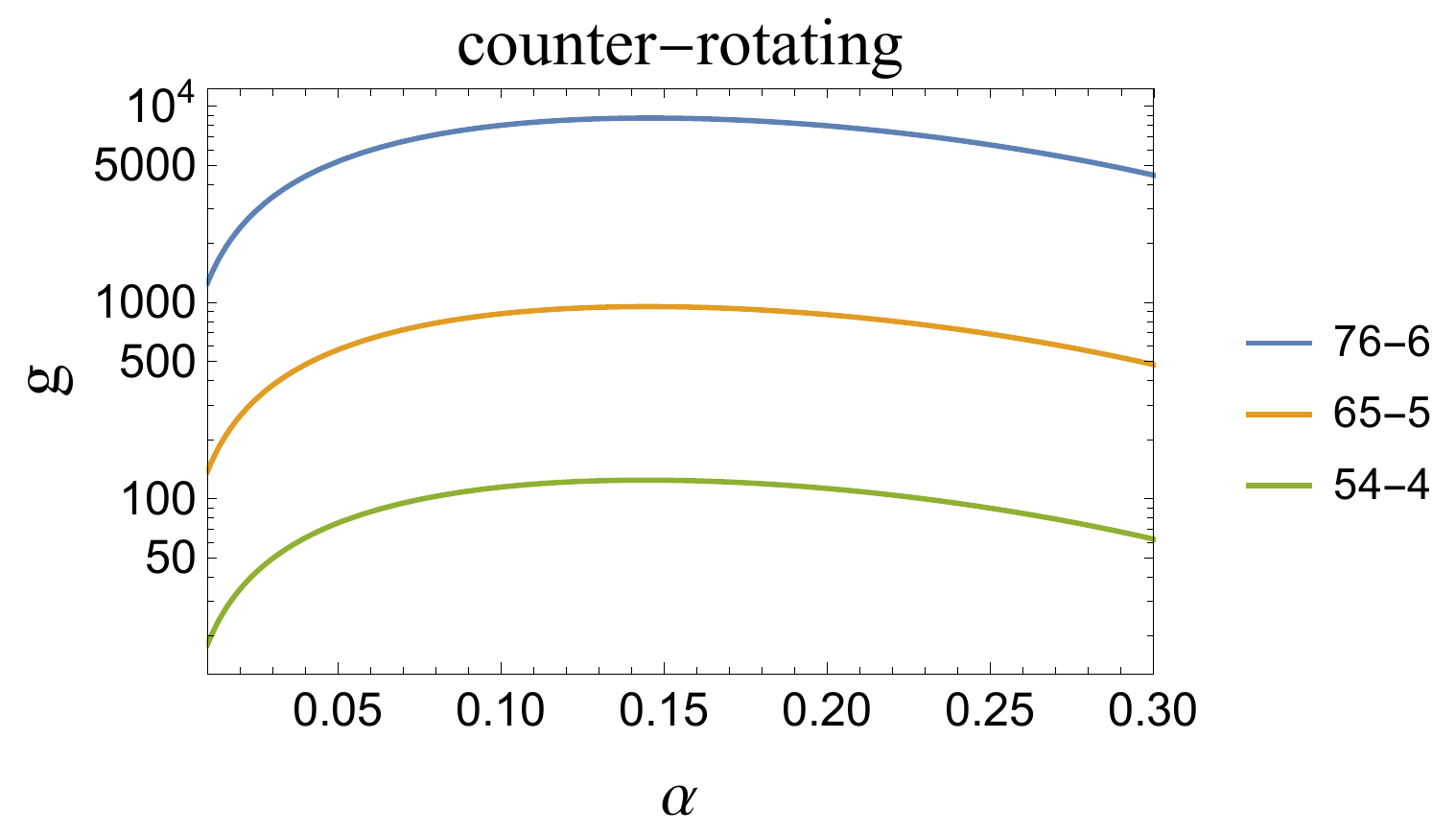}
\end{minipage}
\end{tabular}
	\caption{Non-linearity $g$ of the transition from $\ket{211}$ for the co-rotating case (left) and the counter-rotating case (right) with $q=1$. Each line corresponds to a different transition destination.}
	\label{figgnl}
\end{figure}

For the system that we are considering, the conservation of angular momentum reads
\begin{equation}\label{AngBal}
\frac{\mathrm{d}}{\mathrm{d}t}\left(J_{\mathrm{binary}}\pm J_{\mathrm{c}}\right)=-{\cal T}_{\mathrm{GW}}~, 
\end{equation}
where the right hand side is the torque caused by the radiation reaction 
due to the gravitational wave emission. We choose the signs of $J_{\rm binary}$ and 
${\cal T}_{GW}$ to be positive, irrespectively of the direction of the orbital rotation. 
The upper (lower) sign corresponds to the co-rotating (counter-rotating) case. 
The angular momentum of the cloud can be written as
\begin{equation}
J_{\mathrm{c}}=J_{\mathrm{c},0}\left[m_{1}|c_{1}(t)|^2+m_{2}|c_{2}(t)|^2\right]~, 
\end{equation}
where we defined $J_{c,0}$ so that the initial angular momentum of the cloud 
is given by $m_1 J_{c,0}$. 
When we consider the transition between the modes with a small decay rate, 
we can ignore the time variation of the BH spin during the transition. 
Then, we have the time evolution of the orbital angular velocity around the resonance frequency as~\cite{Baumann:2019ztm}
\begin{equation}\label{OeqBR}
\frac{\mathrm{d}\Omega}{\mathrm{d}t}=\gamma\pm3R_{J}\Omega_{0}\frac{\mathrm{d}}{\mathrm{d}t}\left[m_{1}|c_{1}|^2+m_{2}|c_{2}|^2\right]~,
\end{equation}
where
\begin{equation}
R_{J}=\frac{J_{c,0}}{M^2}\frac{(1+q)^{1/3}}{q}(M\Omega_{0})^{1/3}~.
\end{equation}
Integrating Eq.(\ref{OeqBR}) with the initial condition $\Omega(t)\to\Omega_{0}+\gamma t$ at $t\to-\infty$, we have
\begin{equation}
\Omega(t)=\Omega_{0}+\gamma t\pm3R_{J}\Omega_{0}\left[m_{1}(|c_{1}|^2-1)+m_{2}|c_{2}|^2\right]~.
\end{equation}
Therefore, from Eq.(\ref{H}), the level transition with backreaction 
is described by the following non-linear Hamiltonian
\begin{equation}\label{nH}
\mathcal{H}=\left(\begin{array}{cc}
\theta(t) & \eta \\
\eta & -\theta(t)
\end{array}\right)~, 
\end{equation}
\begin{equation}\label{th}
\theta(t)=\pm\frac{\Delta m}{2}\gamma t +\frac{3|\Delta m|^2}{2}R_{J}\Omega_{0}|c_{2}|^2~,
\end{equation}
where we used $|c_{1}|^2+|c_{2}|^2=1$. 
We now introduce a parameter which represents 
the strength of the non-linearity, 
\begin{equation}\label{gdef}
g\equiv\frac{3|\Delta m|^2 R_{J}\Omega_{0}}{2\eta}~.
\end{equation} 
As an example, we show the value of the non-linearity $g$ for some transitions with $q=1$ in Fig.~\ref{figgnl}. 
When higher-multipole moments are concerned, 
the effect of the non-linearity can be especially large. 
For this reason, it is important to investigate the transition including this backreaction.

As shown in Ref.~\cite{Baumann:2019ztm}, there are two cases in the way how 
the backreaction affects the evolution of the orbital motion. 
They are classified by the relative sign of the terms in Eq.\eqref{th}. 
Case(1) is the case where the relative sign is positive.
This occurs for the "co-rotating and $\Delta m>0$" case or 
for the "counter-rotating and $\Delta m<0$" case. 
In case(1), the cloud receives the angular momentum from the orbital motion of the binary, and the orbit sinks faster. 
Hence, the binary inspiral passes through the resonance band more quickly, 
which makes the persistent rate larger. 
Case(2) is the case where the relative sign is negative.
This occurs for the "co-rotating and $\Delta m<0$" case or 
for the "counter-rotating and $\Delta m>0$" case. 
In case(2), the cloud gives the angular momentum to the orbital motion, 
and the orbital evolution tends to stagnate. 
Hence, the binary inspiral passes through the resonance band more slowly, 
which makes the transition more adiabatic and the persistent rate smaller.

The first transition during the binary inspiral for the counter-rotating case 
belongs to the case(1). 
On the other hand, the first transition between hyperfine split for the 
co-rotating case seems to belong to the case(2). 
We will discuss this point in detail in Sec.~\ref{BRh}.

\subsection{Backreaction to the hyperfine split through the change in the geometry}\label{BRh}
We consider a cloud that occupies only the level with the maximum superradiant growth rate. 
Usually, we can ignore the backreaction to the geometry when we discuss the evolution of the cloud because of the low energy density of the cloud~\cite{Brito:2014wla}. 
However, focusing on the very small hyperfine split, which is proportional to the angular momentum, 
we may not be able to ignore the backreaction due to the change of the angular momentum of the saturated cloud, because it is expected to be comparable to that of the central BH. 
To take this effect into account, we model the frequency shift related to the coupling with the background angular momentum, which corresponds to the fifth term in Eq.\eqref{Ene}, as 
\begin{equation}
\delta\omega^{h}(t)=\mu\frac{2m\alpha^5}{n^3 l(l+1/2)(l+1)}\frac{J}{M^2}\left(1+{x}\frac{J_{\mathrm{c}}(t)}{J}\right)~.
\end{equation}
We parameterized the efficiency of the frequency shift due to the angular momentum of the cloud itself by ${x}$, which should be $\mathcal{O}(1)$. In fact, we evaluate the value of $x$ in Appendix.~\ref{AppB} for the flat background
\footnote{If we consider the maximally grown cloud, it will have a larger angular momentum than that of the BH for $\alpha\ll 1$. Hence, the frequency shift due to the angular momentum of the cloud becomes dominant, and the resonance frequency deviates from the one calculated from Eq.\eqref{Ene}. This means that we should consider the deviation from the Kerr metric when we need the precise value of the resonance frequency. However, it would not change our results qualitatively, and we leave this problem as a future work.}
. 

\begin{figure}[t]
	\centering
	\includegraphics[keepaspectratio,scale=0.6]{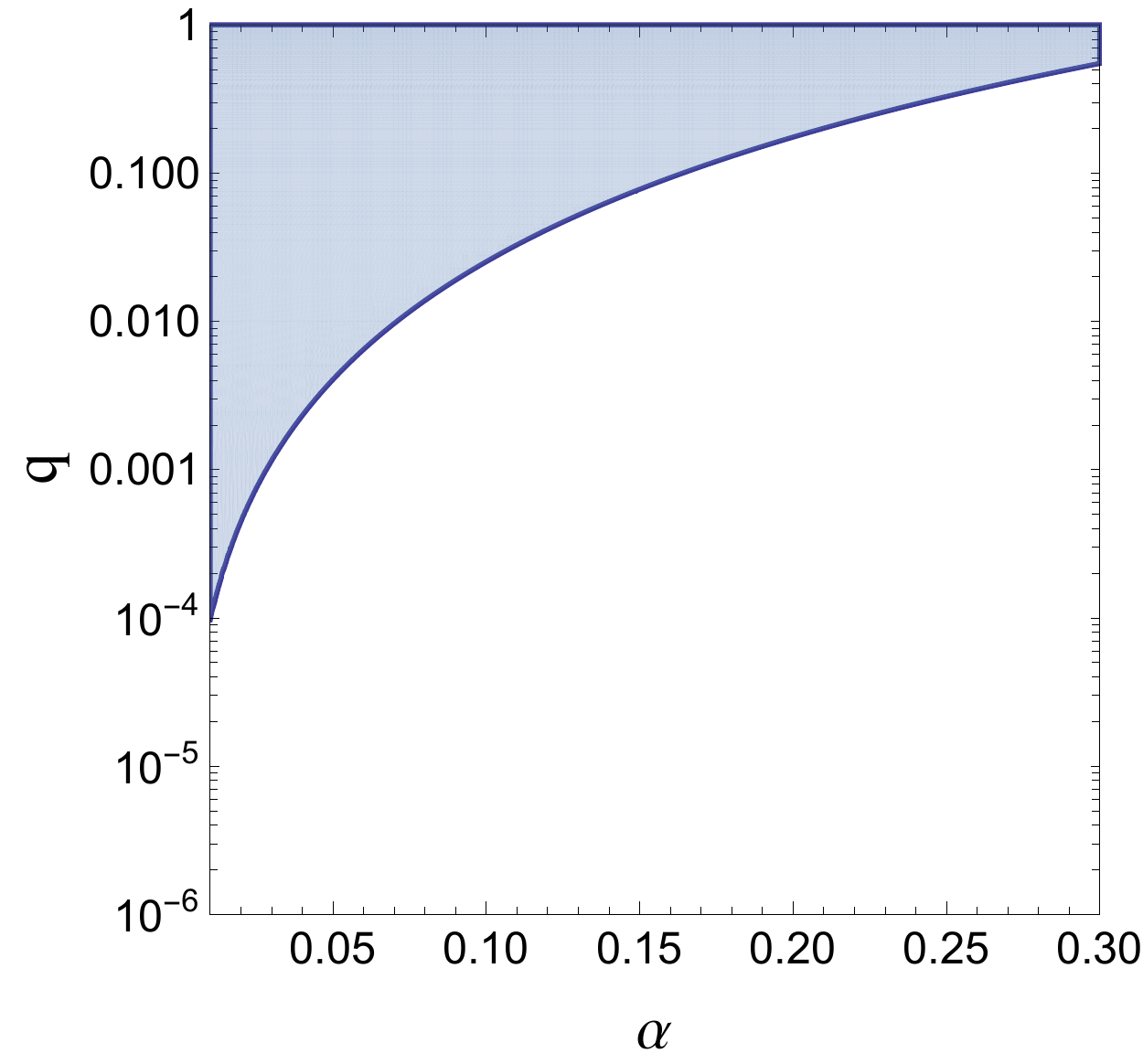}
	\caption{Parameter region that satisfies $3R_{J}<xJ_{\mathrm{c},0}/(J+xm_{1}J_{\mathrm{c},0})$ for $\ket{211}\to\ket{21\mbox{-1}}$ with $J/M^2=a_{\mathrm{crit}}/M,J_{\mathrm{c},0}/M^2=1-J/M^2,x=7/32$. In the shaded region, this hyperfine transition belongs to the case(1) due to the shrink of the level gap according to the angular momentum transfer.}
	\label{figHyperfine}
\end{figure}

In this section our interest is in the first transition for the co-rotating case, $\ket{211}\to\ket{21\mbox{-1}}$, {\it i.e.},
$\Delta m<0$. 
So far, we have considered the resonance frequency $\Omega_0$ to be a constant. However, in this case, since the level gap varies with time, we consider the resonance frequency to be time-dependent
\footnote{It is only in this section that the resonance frequency to be time-dependent.}.   
We define the initial resonance frequency by
\begin{equation}
\Omega_{0}^{(-\infty)}=\mu\frac{2\alpha^5}{n^3 l(l+1/2)(l+1)}\frac{J}{M^2}\left(1+x\frac{J_{\mathrm{c}}(-\infty)}{J}\right)~,
\end{equation}
where $J_{\mathrm{c}}(-\infty)=m_{1}J_{\mathrm{c},0}$. Then, the time-dependent resonant frequency becomes
\begin{equation}
\Omega_{0}(t)=\Omega_{0}^{(-\infty)}+\frac{xJ_{\mathrm{c},0}}{J+xm_{1}J_{\mathrm{c},0}}\Omega_{0}^{(-\infty)}\left[m_{1}(|c_{1}(t)|^2-1)+m_{2}|c_{2}(t)|^2\right]~.
\end{equation}
This implies that the resonance band shrinks as the angular momentum is transferred from 
the cloud to the orbital motion. 
Thus, we have the diagonal term of the Hamiltonian (\ref{H}) that describes the transition between the levels separated by the hyperfine split as
\begin{align}
\theta^{h}(t)&=-\frac{|\Delta m|}{2}\left(\Omega(t)-\Omega_{0}(t)\right) \notag \\
&=-\frac{|\Delta m|}{2}\gamma t +\frac{|\Delta m|^2}{2}\Omega_{0}^{(-\infty)}\left(3R_{J}-\frac{xJ_{\mathrm{c},0}}{J+xm_{1}J_{\mathrm{c},0}}\right)|c_{2}|^2~.
\end{align}
When we ignore the backreaction on the geometry, {\it i.e.}, for $x=0$, this transition belongs to the case(2). 
However, if $3R_{J}<xJ_{\mathrm{c},0}/(J+xm_{1}J_{\mathrm{c},0})$, 
the relative sign reverses and this transition belongs to the case(1). 
In Fig.~\ref{figHyperfine}, we show the parameter region that satisfies $3R_{J}<xJ_{\mathrm{c},0}/(J+xm_{1}J_{\mathrm{c},0})$. 
For example, in the case of equal mass binaries, 
this condition is satisfied and the transition effectively belongs to the case(1). 
In other words, since the effect of the shrinking the resonance frequency works more efficiently than the effect of slowing down the evolution of the orbit, the resonance band is passed through quickly and the persistent rate becomes larger. 

In this hyperfine transition, the non-linearity $g$ given in Eq.(\ref{gdef}) should be replaced with the effective one, 
\begin{equation}
g_{\mathrm{eff}}\equiv\frac{|\Delta m|^2}{2\eta} \left|3R_{J}-\frac{xJ_{\mathrm{c},0}}{J+xm_{1}J_{\mathrm{c},0}}\right|\Omega_{0}^{(-\infty)}~.
\end{equation}
In particular, for the transition $\ket{211}\to\ket{21\mbox{-1}}$ with $3R_{J}<xJ_{\mathrm{c},0}/(J+xm_{1}J_{\mathrm{c},0})$, 
the effective non-linearity can be approximately evaluated as
\begin{equation}
g_{\mathrm{eff}}\simeq 3\times10^3\frac{1+q}{q}\left(\frac{0.1}{\alpha}\right)^3\frac{xJ_{{\rm c},0}/M^2}{(J/M^2+xJ_{{\rm c},0}/M^2)^2}~.
\end{equation}
Therefore, even for equal mass binaries, 
this transition is in the strongly non-linear regime. 
This means that the persistent rate is large and most of the axions remain in the state $\ket{211}$. 
The resonances that the binary experiences after the first hyperfine resonance are 
the ones associated with the change of $n$ and $\Delta m>0$. 
Hence, they belong to the case(1). 
As a result, when we consider the binary system with $q=\mathcal{O}(1)$, almost all transitions from $\ket{211}$ belong to the case(1) 
and are in the strongly non-linear regime.

\subsection{Persistent rate for the strongly non-linear regime}
Focusing on the binary system with $q=\mathcal{O}(1)$, we investigate the level transition in the strongly non-linear regime $g\gg1$ for the case(1). 
In particular, we derive the self-consistent equation for the persistent rate 
by referring to Ref.~\cite{PhysRevA.66.023404}.

\begin{figure}[t]
 \begin{tabular}{cc}
 \begin{minipage}[t]{0.5\hsize}
	\centering
	\includegraphics[keepaspectratio,scale=0.4]{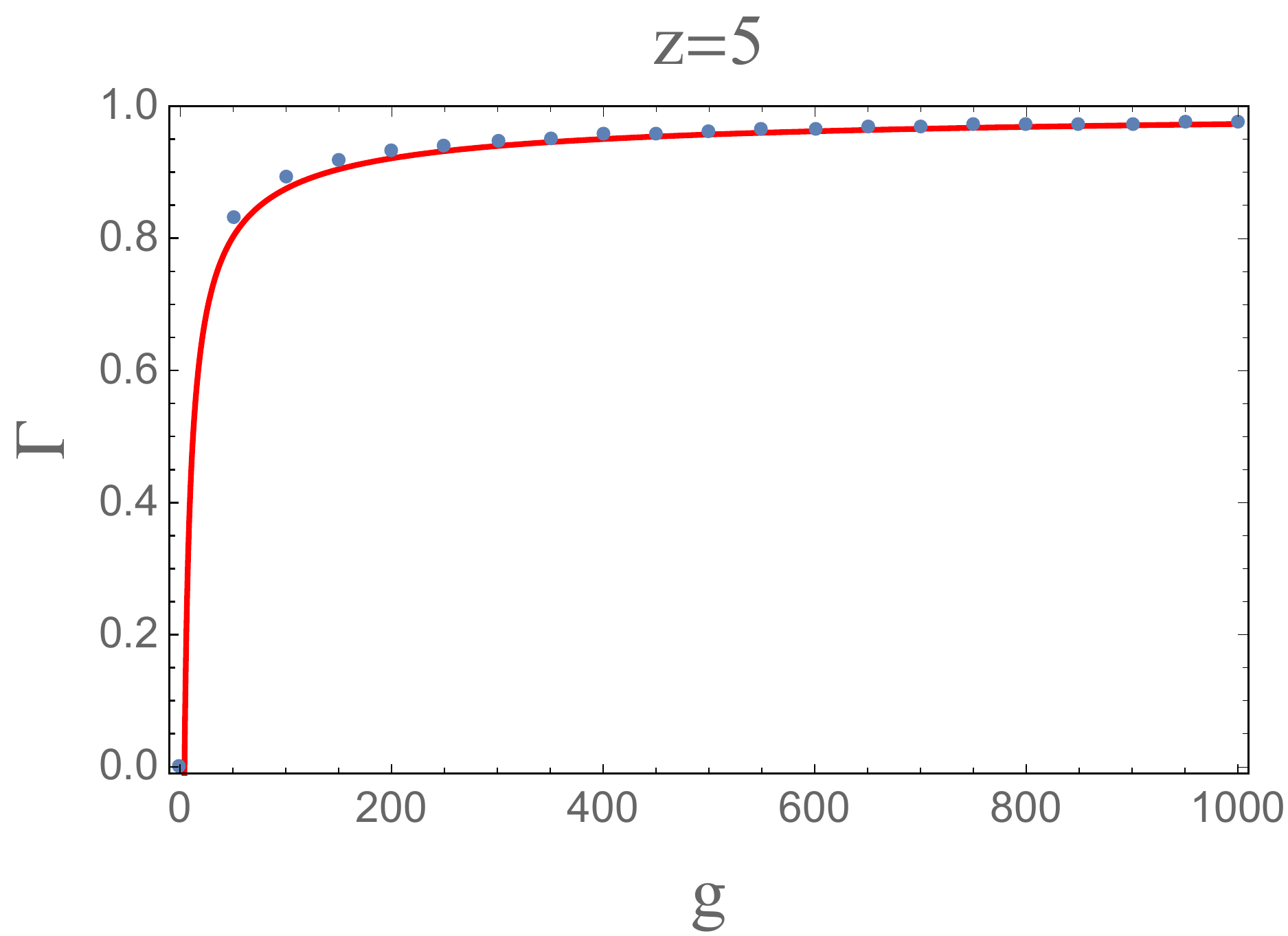}
\end{minipage} &
\begin{minipage}[t]{0.5\hsize}
        \centering
	\includegraphics[keepaspectratio,scale=0.4]{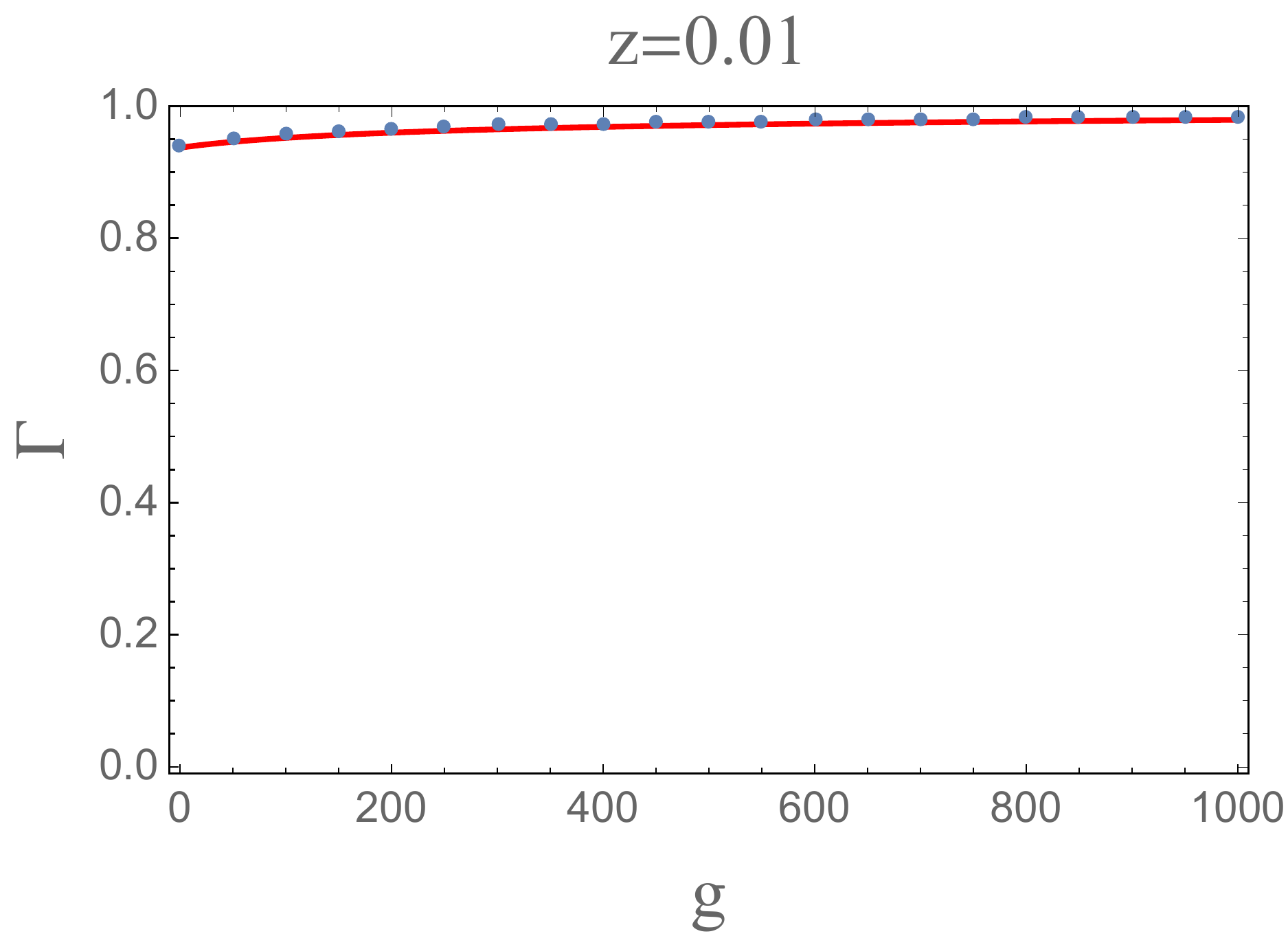}
\end{minipage}
\end{tabular}
	\caption{Persistent rate $\Gamma$ including backreaction whose strength is characterized by the non-linearity $g$ with $z=5$ (left) and $z=0.1$ (right). The red solid line shows the solution of the self-consistent equation (\ref{PRwBR}) for strong non-linear regime. The blue dots show the numerical results obtained by solving differential equations described by the non-linear Hamiltonian (\ref{nH}).}
	\label{figPR}
\end{figure}

For the case(1), $\theta(t)$ in the diagonal elements of the Hamiltonian Eq.\eqref{nH} is
\begin{equation}
\theta(t)=\frac{|\Delta m|}{2}\gamma t +\frac{3|\Delta m|^2}{2}R_{J}\Omega_{0}|c_{2}|^2~.
\end{equation}
First, we transform the coefficients as 
\begin{align}
c_{1}=\bar{c}_{1}e^{-i\int_{0}^{t}dt'\theta}~, \\
c_{2}=\bar{c}_{2}e^{+i\int_{0}^{t}dt'\theta}~.
\end{align}
By this transformation, the diagonal elements in Eq.\eqref{nH} vanish, and 
the time evolution of $\bar{c}_{2}$ is determined by
\begin{equation}
\frac{\mathrm{d}\bar{c}_{2}}{\mathrm{d}t}=\frac{\eta}{i}\bar{c}_{1}e^{-2i\int_{0}^{t}dt'\theta}~.
\end{equation}
Integrating this equation with the initial condition $\bar{c}_{2}(-\infty)=0$, we formally obtain
\begin{equation}
\bar{c}_{2}(t)=\frac{\eta}{i}\int_{-\infty}^{t}dt'\bar{c}_{1}(t')\exp\left[-2i\int_{0}^{t'}dt''\theta(t'')\right]~.
\end{equation}
Because $g\gg 1$ and the exponent in the above expression shows rapid oscillations, 
we evaluate the above integral using the stationary phase approximation. 
We denote the second derivative coefficient at the stationary point $t_{0}$ by 
\begin{align}\label{gammab}
|\Delta m| \bar{\gamma}&\equiv\left.\frac{\mathrm{d^2}}{\mathrm{d}t^2}\left(2\int_{0}^{t}dt'\theta(t')\right)\right|_{t_{0}} \notag \\
&=|\Delta m| \gamma+3|\Delta m|^2R_{J}\Omega_{0}\left.\frac{\mathrm{d}|c_{2}|^2}{\mathrm{d}t}\right|_{t_{0}}~. 
\end{align}
Then, we have an expression for the particle number as
\begin{align}
|c_{2}(t)|^2=&\eta^2|\bar{c}_{1}(t_{0})|^2\left(\int_{-\infty}^{t}dt'\exp\left[\frac{i}{2}|\Delta m|\bar{\gamma}(t'-t_{0})^2\right]\right)
\cr  & \qquad\qquad\qquad \times \left(\int_{-\infty}^{t}dt'\exp\left[-\frac{i}{2}|\Delta m|\bar{\gamma}(t'-t_{0})^2\right]\right)~.
\end{align}
From this expression, we get the form of Fresnel integral and can calculate the following quantities;
\begin{equation}\label{dc2t0}
\left.\frac{\mathrm{d}|c_{2}|^2}{\mathrm{d}t}\right|_{t_{0}}=\eta^2|\bar{c}_{1}(t_{0})|^2\sqrt{\frac{\pi}{|\Delta m|\bar{\gamma}}}~,
\end{equation}
\begin{equation}\label{c2inf}
|c_{2}(+\infty)|^2=\eta^2|\bar{c}_{1}(t_{0})|^2\frac{2\pi}{|\Delta m|\bar{\gamma}}~.
\end{equation}
Using the above results, we evaluate the persistent rate $\Gamma=|c_{1}(+\infty)|^2=1-|c_{2}(+\infty)|^2$ self-consistently. Here, $\Gamma$ is nearly one because of the strong non-linearity, so we approximate $|\bar{c}_{1}(t_{0})|^2\simeq 1$.
From Eq.(\ref{c2inf}), we get
\begin{equation}
\frac{1}{1-\Gamma}=\frac{|\Delta m|\bar{\gamma}}{2\pi\eta^2}~.
\end{equation}
Substituting Eq.(\ref{gammab}) into this equation and using Eq.(\ref{dc2t0}), finally we obtain a closed form of the equation for $\Gamma$ as
\begin{equation}\label{PRwBR}
\frac{1}{1-\Gamma}=\frac{1}{2\pi z}+\frac{g}{\sqrt{2}\pi}\sqrt{1-\Gamma}~,
\end{equation}
where the defenitions of $z$ and $g$ are given by Eq.\eqref{zdef} and Eq.\eqref{gdef}, respectively. Solving this equation numerically, we can estimate the persistent rate including the backreaction without solving non-linear differential equations. 
We compare the solution of Eq.(\ref{PRwBR}) with the numerical result for the persistent rate obtained by solving Eq.(\ref{nH}) directly in Fig.~\ref{figPR}. 
We can see that Eq.(\ref{PRwBR}) gives a good approximation of the persistent rate with backreaction for $g\gg 1$.

\begin{figure}[t]
 \begin{tabular}{cc}
 \begin{minipage}[t]{0.5\hsize}
	\centering
	\includegraphics[keepaspectratio,scale=0.6]{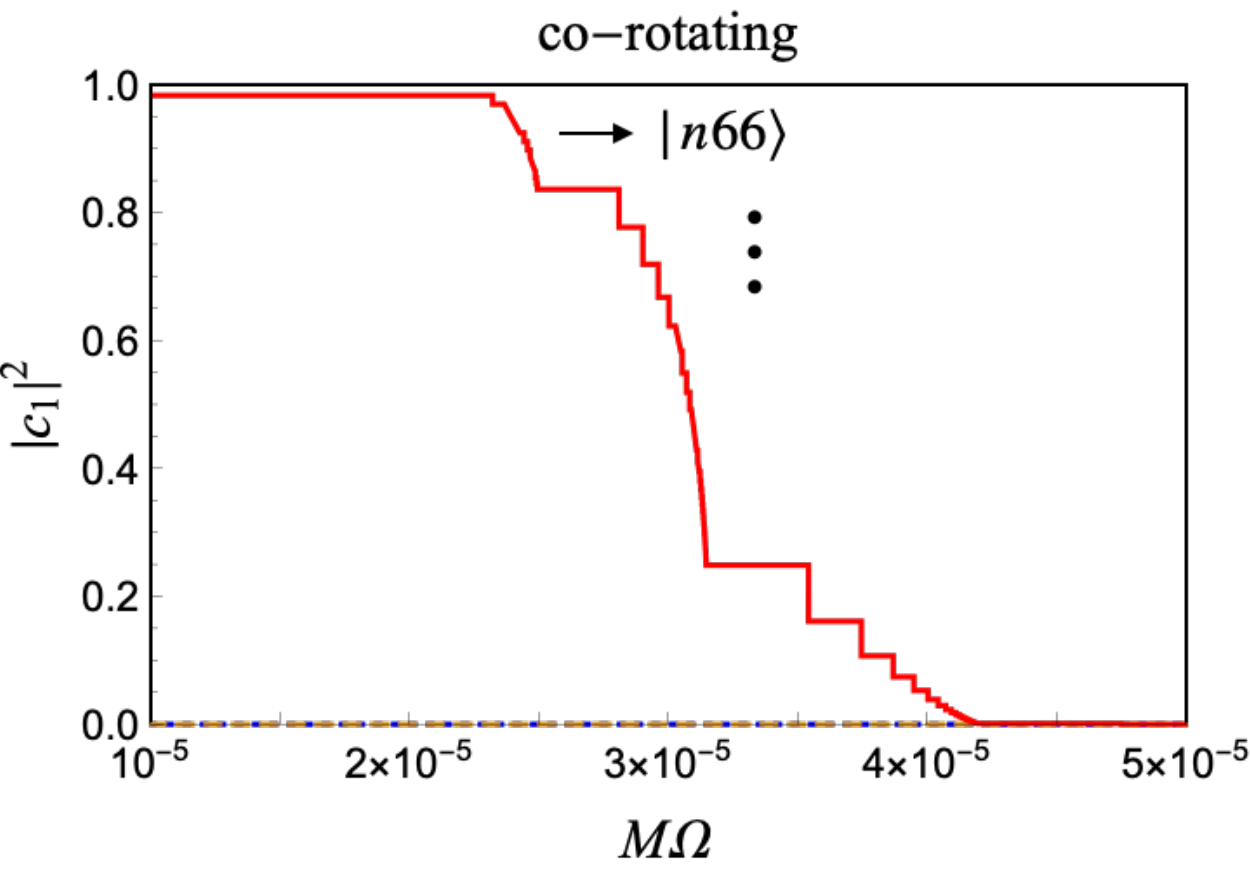}
\end{minipage} &
\begin{minipage}[t]{0.5\hsize}
        \centering
	\includegraphics[keepaspectratio,scale=0.6]{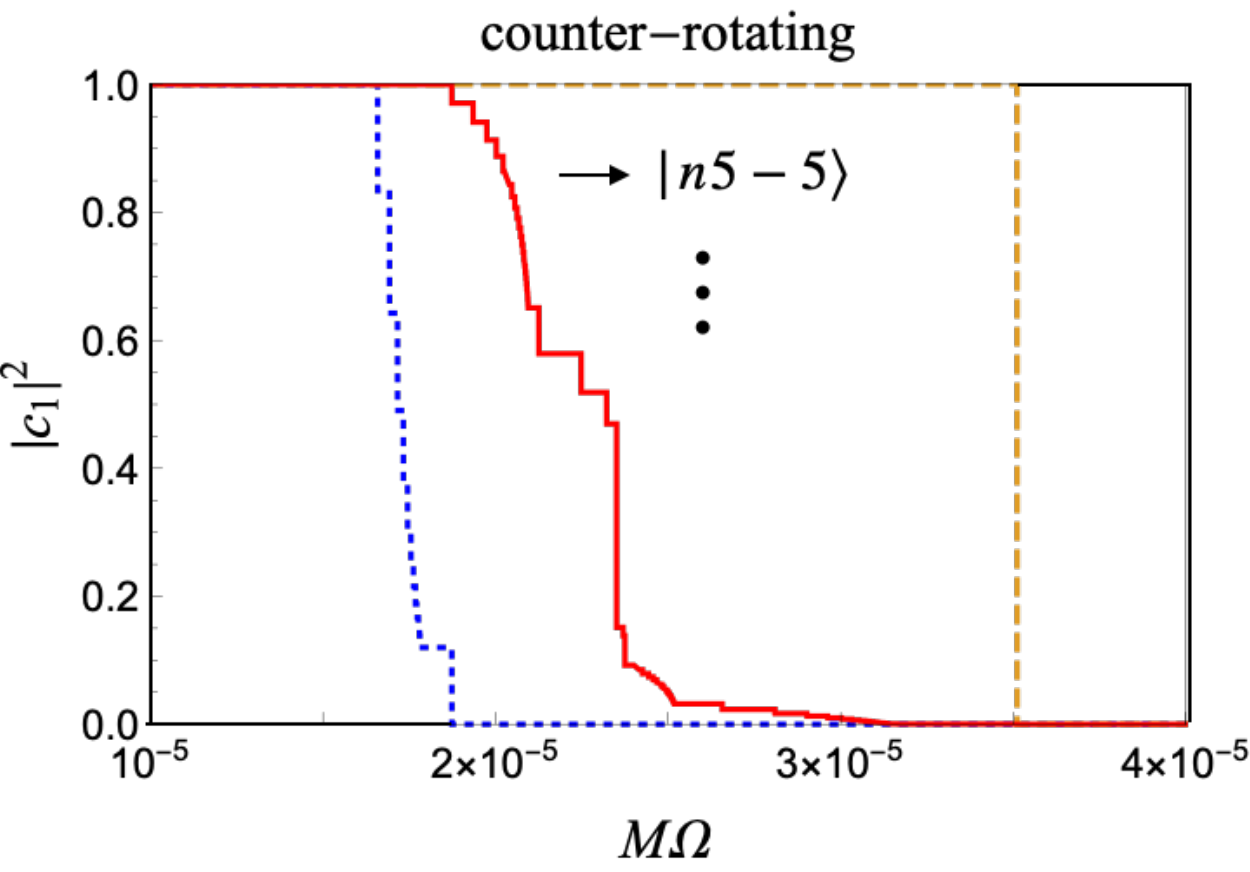}
\end{minipage}
\end{tabular}
	\caption{Evolution of the normalized particle number $|c_{1}|^2$ occupying $\ket{211}$,  including the backreaction for the co-rotating case (left) and the co-rotating case (right) with $\alpha=0.1,q=1$. The red solid curve shows the transition taking into account the backreaction and the higher-multipole moments. The blue and yellow curves are the same as in Fig.~\ref{figHMtr}. Note that, for the co-rotating case, since the transition between hyperfine split is passed through with little transition, the blue and yellow lines are almost zero in this range. }
	\label{figBRtr}
\end{figure}

Using this formula, we investigate how the evolution of the particle number that we have shown in Fig.~\ref{figHMtr} is modified. 
We show the results obtained by replacing the persistent rate Eq.(\ref{lPR}) with the one evaluated using Eq.(\ref{PRwBR}) in Fig.~\ref{figBRtr}
\footnote{In the case(1) with strong non-linearity, transition becomes non-adiabatic and particle number changes with oscillations. 
Since the orbital frequency also oscillates at the transition~\cite{Baumann:2019ztm}, 
strictly speaking, it may not be appropriate to treat the orbital frequency as a time coordinate.}
\footnote{In this figure, we use the persisitent rate obtained by solving Eq.(\ref{nH}) directly for transitions with $g<10$.}
.
For the counter-rotating case, the qualitative results do not change. By contrast, for the co-rotating case, the axion cloud passes through the resonance corresponding to the transition $\ket{211}\to\ket{21\mbox{-1}}$ with little transition. 
Later, the axions are transferred to higher levels as in the counter-rotating case.

When the mass ratio is small, {\it i.e.}, $q\ll 1$, the first transition for the co-rotating case $\ket{211}\to\ket{21\mbox{-1}}$ belongs to the case(2), even if we take into account the reduction of the hyperfine split due to the angular momentum loss of the cloud. 
In this case, we cannot apply the method we introduced in this section straightforwardly because of the large change in the number of particles occupying the initial state. 
In addition, for such a transition,  since the decay rate of the transition destination 
is large compared with the timescale of the evolution of the binary orbital frequency, 
we may not be able to treat the level transition as usual. 
We leave a study on the extreme mass ratio inspiral case as future work.

\section{Axion emission}\label{Sec4}
So far, we discussed the first transition destination and found that the axions in $\ket{211}$ are transferred to higher levels for both directions of the orbital rotation, at least for $q=\mathcal{O}(1)$. 
At this point, since the decay (growth) rates of the transition destinations are generally very small compared with the timescale of the binary evolution, 
axions are not reabsorbed by the BH before the merger. 
However, axions once excited to higher levels can be easily further excited by the tidal interaction to an unbounded state. 
In this section, we evaluate the particle number flux to the infinity.

\subsection{Particle number flux}
We evaluate the particle number flux perturbatively. 
We solve the following approximate equations,
\begin{align}
\left(i\partial_t+\frac{1}{2\mu}\nabla^2+\frac{\alpha}{r}\right)\psi^b&=0~, \\
\left(i\partial_t+\frac{1}{2\mu}\nabla^2+\frac{\alpha}{r}\right)\psi^r&=V_{\ast}\psi^b~,
\end{align}
where 
\begin{equation}
\psi^{b}(x)=e^{-i(\omega_{0}-\mu) t}R_{n_{0}l_{0}}(r)Y_{l_{0}m_{0}}(\theta,\varphi)
\end{equation}
 is the background bound state solution and $\psi^{r}$ is the radiative part. Here, we normalize the number of particles that occupy the background state as
 \begin{equation}
     \int d^3 x|\psi^{b}(x)|^2=1~.
 \end{equation}
 We construct the solution for the radiative part perturbatively using Green's function as
 \begin{equation}
 \psi^{r}(x)=\int d^{4}x' G(x,x')V_{\ast}\psi^{b}(x')~.
 \end{equation}
 Here, the Green's function is defined to satisfy
 \begin{equation}
 \left(i\partial_t+\frac{1}{2\mu}\nabla^2+\frac{\alpha}{r}\right)G(x,x')=\delta^4(x-x')~.
 \end{equation}
 We can decompose the Green's function as
 \begin{equation}
 G(x,x')=\sum_{lm}\int\frac{d\omega}{2\pi}e^{-i(\omega-\mu)(t-t')}Y_{lm}(\theta,\varphi)Y_{lm}^{\ast}(\theta',\varphi')G_{l}^{\omega}(r,r') ~.
 \end{equation}
 The radial part of the Green's function $G_{l}^{\omega}(r,r')$ can be constructed using the Coulomb wavefunction $R_{kl}(r)$ that satisfies
 \begin{equation}
 \frac{1}{2\mu}\left(\frac{1}{r^2} \frac{{\rm d}}{{\rm d}r} r^2 \frac{{\rm d}}{{\rm d}r}   -\frac{l(l+1)}{r^2}+k^2+\frac{2\mu\alpha}{r}\right)R_{kl}(r)=0~,
 \end{equation}
 where $k=\sqrt{2\mu(\omega-\mu)}$. We have
 \begin{gather}
G_{l}^{\omega}(r,r')=\frac{2\mu}{W_{kl}}\left(R^{0}_{kl}(r)R^{+}_{kl}(r')\theta(r'-r)+R^{0}_{kl}(r')R^{+}_{kl}(r)\theta(r-r')\right)~, \end{gather}
with
\begin{gather}
W_{kl}=r^2\left(R^{0}_{kl}\frac{{\rm d}R^{+}_{kl}}{{\rm d}r}-R^{+}_{kl}\frac{{\rm d} R^{0}_{kl}}{{\rm d}r}\right)~.
\end{gather}
$R_{kl}^{0}$ is regular at the origin and $R_{kl}^{+}(r)$ is the outgoing wave 
at infinity (see Appendix~\ref{mathfunc} for the explicit form). 
Now, assuming the constant orbital angular velocity of the binary, {\it i.e.}, $\Phi_{\ast}(t)=\pm\Omega t$, we consider the tidal field given by 
\begin{equation}
V_{\ast}=\sum_{l_{\ast}m_{\ast}}V_{\ast,l_{\ast}m_{\ast}}e^{\mp im_{\ast}\Omega t}~.
\end{equation}
Then, we obtain the perturbative solution, 
\begin{align}
\psi^{r}&=\sum_{lm}\sum_{l_{\ast}m_{\ast}}\int d^{4}x' \int \frac{d\omega}{2\pi}e^{-i(\omega-\mu)(t-t')}Y_{lm}(\theta,\varphi)Y_{lm}^{\ast}(\theta',\varphi') \notag \\
&\qquad \qquad \times  V_{\ast,l_{\ast}m_{\ast}}(r',\theta',\varphi')e^{\mp im_{\ast}\Omega t'}G_{l}^{\omega}(r,r')e^{-i(\omega_{0}-\mu)t'}R_{n_{0}l_{0}}(r')Y_{l_{0}m_{0}}(\theta',\varphi') \notag \\
&\to-\sum_{lm}\sum_{l_{\ast}m_{\ast}}\int d\omega\ \delta(\omega-\omega_{0}\mp m_{\ast}\Omega)e^{-i(\omega-\mu)t}\frac{e^{ikr}}{r}Y_{lm}(\theta,\varphi)\frac{2\mu}{k} \notag \\
&\qquad \qquad \times\int dr'r'{}^2 d\theta' \sin\theta' d\varphi' R^{0}_{kl}(r')Y^{\ast}_{lm}(\theta',\varphi')V_{\ast,l_{\ast}m_{\ast}}(r',\theta',\varphi')R_{n_0l_0}(r')Y_{l_0m_0}(\theta',\varphi')\,,
\end{align}
where we take the limit $r\to\infty$ and choose 
the normalization of $R^{0}_{kl}$ as given 
in Appendix~\ref{mathfunc}. 
Therefore, we can calculate the normalized particle number flux to infinity as
\begin{align}
F&=\int d\theta\sin\theta d\varphi \frac{r^2}{2\mu i}\left(\psi^{r\ast}\partial_{r}\psi^{r}-\psi^{r}\partial_{r}\psi^{r\ast}\right) \notag \\
&=\sum_{lm}\sum_{l_{\ast}m_{\ast}}\frac{4\mu}{\sqrt{2\mu((\omega_0\pm m_{\ast}\Omega)-\mu)}}\left|\int d^{3}x' R^{0}_{kl}Y^{\ast}_{lm}V_{\ast,l_{\ast}m_{\ast}}R_{n_0l_0}Y_{l_0m_0}\right|^2~.
\end{align}

\subsection{Results}
Here, we show how efficiently this emission works in comparison with the timescale of the binary evolution;
\begin{equation}
T_{\mathrm{binary}}\equiv\frac{\Omega}{\gamma}=\frac{5}{96}M\frac{(1+q)^{1/3}}{q}(M\Omega)^{-8/3}~.
\end{equation}
First, for simplicity, we give an analytic estimate of the flux from the cloud composed of $\ket{211}$ state in the limit $k\to 0$. This limit corresponds to the case where the binary orbital frequency is just at the threshold for the excitation to an unbound state. 
To estimate the flux, we consider the component that gives the maximum amplitude of the tidal field, {\it i.e.}, the flux due to the $(l_{\ast},m_{\ast})=(2,\pm2)$ perturbation. 
Hence, 
\begin{equation}
M\Omega\simeq\frac{1}{|\Delta m|}\frac{\alpha^3}{2n^2}=\frac{\alpha^3}{16}~.
\end{equation}
At this frequency, the timescale of the binary's evolution is given by
\begin{equation}
T_{\mathrm{binary}}\simeq8\times10^{9}\frac{(1+q)^{1/3}}{q}M\left(\frac{0.1}{\alpha}\right)^8~.
\end{equation}
In the limit $k\to 0$, the radiative wave function is given by 
\begin{equation}
R_{kl}^{0}\to\sqrt{k}\sqrt{\pi/r}J_{2l+1}(\sqrt{8 \mu \alpha r})~,
\end{equation}
where $J_{\nu}(x)$ is the Bessel function. 
Then, we can roughly estimate the flux as
\footnote{Here, we consider only the inner part of perturbation in Eq.(\ref{Vast}) and underestimate the actual value. 
However, the flux is still so large that it should not affect the conclusion.}
\begin{align}
F&\simeq2\times10^{-5}\frac{1}{M}\frac{q^2}{(1+q)^2}\left(\frac{\alpha}{0.1}\right)^3 \quad \textrm{(co-rotating)}~, \\
F&\simeq4\times10^{-6}\frac{1}{M}\frac{q^2}{(1+q)^2}\left(\frac{\alpha}{0.1}\right)^3 \quad \textrm{(counter-rotating)}~.
\end{align}
The particle number flux divided by the number of particles contained in the cloud is sufficiently large compared with the timescale of the binary evolution, 
except for the extreme mass ratio inspiral case with $q\ll1$. 
It means that if the orbital frequency reaches the threshold for the transition to an unbound state, bounded axions are all radiated away.

\begin{figure}[t]
 \begin{tabular}{cc}
 \begin{minipage}[t]{0.5\hsize}
	\centering
	\includegraphics[keepaspectratio,scale=0.53]{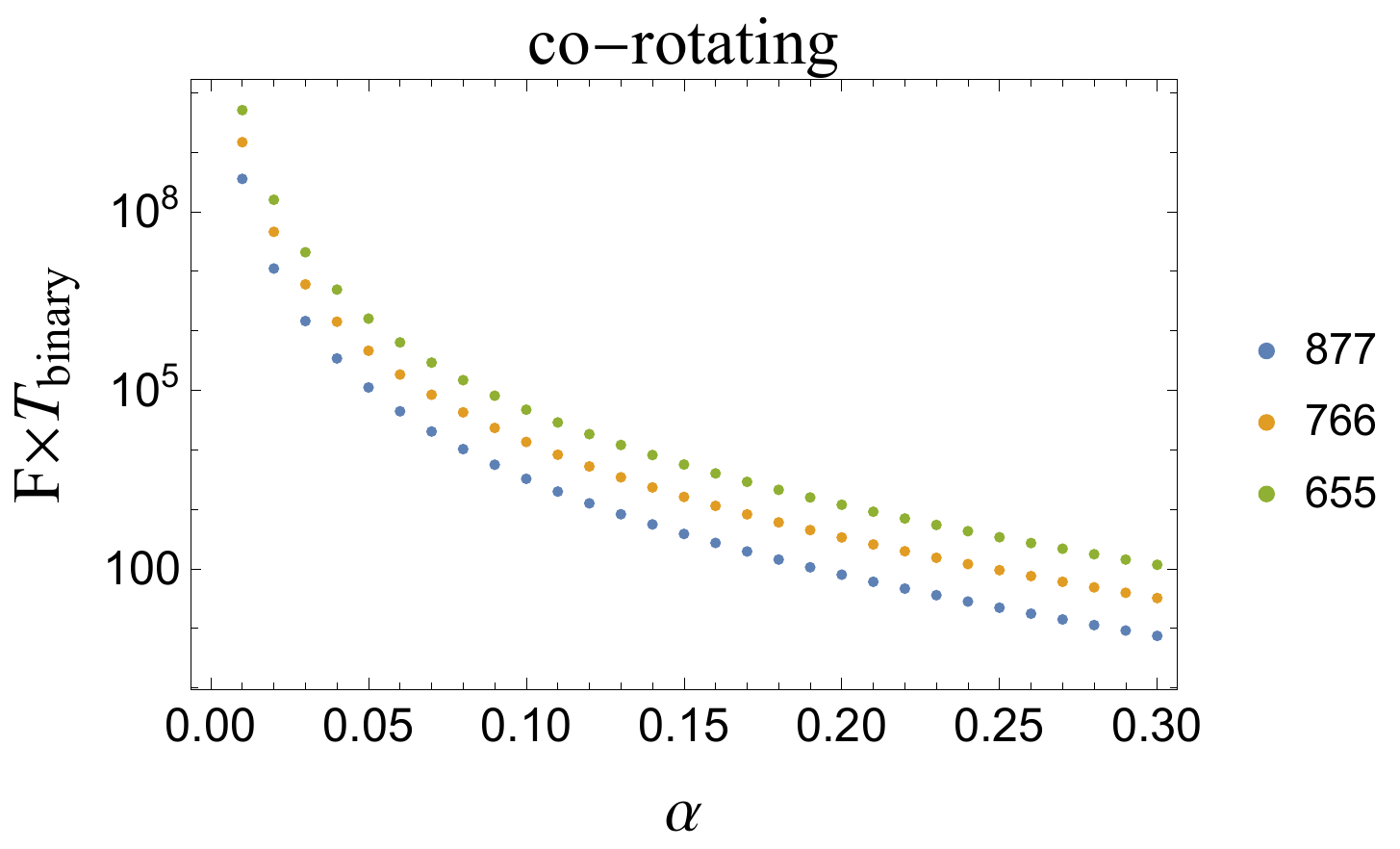}
\end{minipage} &
\begin{minipage}[t]{0.5\hsize}
        \centering
	\includegraphics[keepaspectratio,scale=0.53]{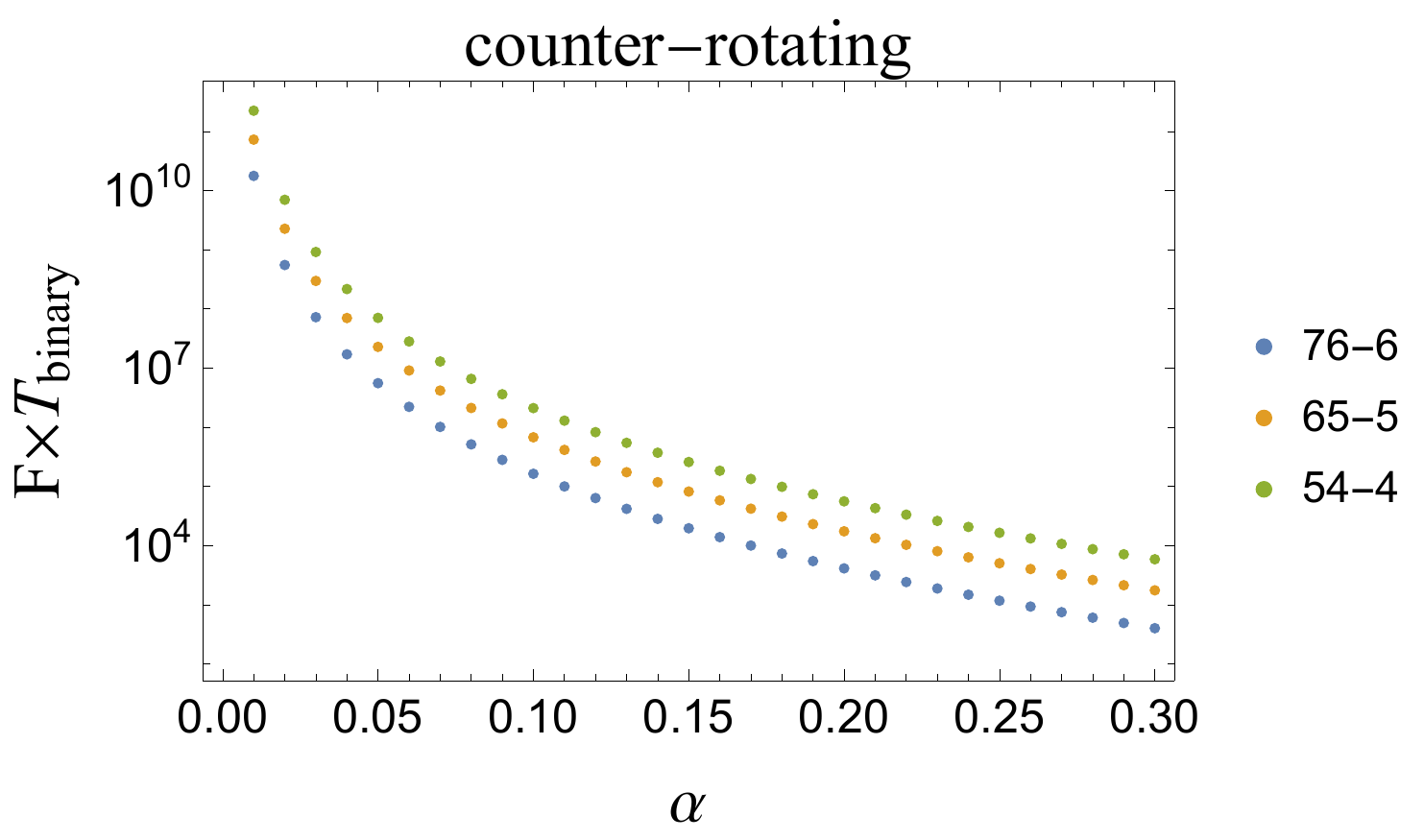}
\end{minipage}
\end{tabular}
	\caption{Comparison between the normalized particle number flux and the timescale of the binary's evolution for both the co-rotating case (left) and the counter-rotating case (right). They are evaluated at the resonance frequency for the transition 
	from $\ket{211}$ to each level for $q=1$.}
	\label{figFlux}
\end{figure}

In practice, as we showed in the previous section, axions are first transferred to the levels with some higher $l$. Hence, We show the numerical results of the comparison between the flux from higher levels and the timescale of the orbital evolution due to radiation reaction in Fig.~\ref{figFlux} for $q=1$. 
The orbital frequency is set to the resonance frequency for the transition from $\ket{211}$ to each level.  As one can see from this figure, axions transferred to higher levels will be radiated away to the infinity rather immediately. 
Therefore, we conclude that the axion cloud disappears just after the first transition, 
through the outward radiation without being reabsorbed by the BH for equal mass binaries.

\section{Summary and Discussion}\label{Sec5}
In this paper, we studied the dynamics of the axion cloud associated with a BH in a binary system during the inspiral phase, for the purpose to give a reliable constraint on the properties of axions from observations of GWs and BHs. 
The presence of the cloud should affect the gravitational waveform, 
especially if it remains until a late inspiral phase close to the merger time. In addition, the level transition of the cloud due to the binary tidal interaction may change the effective forbidden region in the mass and 
spin parameters of BHs observed in binary systems. 
Therefore, in this paper we clarified in which case the cloud disappears, and when it disappears, when and how the process proceeds. 
We mainly focused on equal mass binaries, which are the main target of the ground based gravitational wave detectors, and extended the previous studies in the following manner.

First, we studied level transitions taking into account the higher-multipole moments of the tidal perturbation without the backreaction to the orbital evolution. Though the magnitude of higher-multipole perturbation is 
smaller than that of the leading quadrupole moment, they can be sufficiently strong to make a transition by a non-negligible amount. 
The resonance frequency is roughly inversely proportional to the difference of the azimuthal number between two levels, unless the difference of the energies between two levels is governed by the hyperfine split.  
Therefore, higher-multipole moments can become at work earlier during the inspiral of the binary orbit. We confirmed that this picture is correct and showed the parameter dependence.

Second, we investigated the effects of backreaction due to the angular momentum transfer associated with the level transition, precisely. 
We described the level transition with this backreaction by introducing a non-linear model Hamiltonian and analyzed it. 
In particular, for the transition between two levels gapped by the hyperfine split, which is proportional to the angular momentum of the BH and the cloud, 
we considered the backreaction from the transfer of the angular momentum between the cloud and the orbital motion. 
This effect makes the duration staying at the resonance band shorter and the persistent rate of the initial level larger. 
In addition, we derived the analytic closed formula for the persistent rate in a strongly non-linear regime. Using this formula, we analyzed the transitions with backreaction and found that axions are transferred to the levels with a larger $l$ gradually for both co-rotating and counter-rotating cases.

Finally, we considered the axion emission to the infinity from the cloud induced by the tidal interaction. 
At the first transition, axions are transferred to  higher levels, which do neither decay nor grow rapidly by themselves. 
Hence, axions are not reabsorbed by the BH. 
However, they can be easily excited to a continuous unbound state. 
We calculated the normalized particle number flux and compared it with the timescale of the evolution of the binary orbital frequency. 
As a result, the flux turned out to be large enough to deplete the axion 
clouds completely through the radiation to the infinity. 

From the above results, we can conclude that axion clouds in equal mass binary systems will disappear during the early inspiral phase by the transition to unbound states caused by the tidal perturbation. 
This fact may imply that it is difficult to detect the modulation of the gravitational 
waveform caused by the presence of axion clouds using ground-based gravitational wave detectors. 
On the other hand, because axions are not reabsorbed by the BH, the tidal interaction 
due to the companion in binaries does not affect the evolution of the BH spin. 
It will imply that the current constraints on the possible axion mass from the 
distribution of the mass and the spin of BHs will not be altered, 
even if we use the estimates of these parameters 
derived from gravitational wave observations of binary black holes. 
In this paper, we have concentrated on the theoretical aspects,  
and we leave more careful examination of possible implications to observations, along with the case of extreme mass ratio inspirals, for our future study.

\section*{Acknowledgment}

This work is supported by JSPS Grant-in-Aid for Scientific Research JP17H06358 (and also JP17H06357), as a part of the innovative research area, ``Gravitational wave physics and astronomy: Genesis'', and also by JP20K03928. 
T. Takahashi is supported by the establishment of university fellowships towards the creation of science technology innovation. 


%

\vspace{0.2cm}
\noindent


\let\doi\relax


\appendix

\section{Wavefunction for a hydrogen atom}\label{mathfunc}
In this appendix, we summarize the wavefunction for a hydrogen atom that we use in this paper. We consider the Schr$\mathrm{\ddot{o}}$dinger type equation with the Coulomb potential 
\begin{equation}
\left(i\partial_t+\frac{1}{2\mu}\nabla^2+\frac{\alpha}{r}\right)\psi=0~.
\end{equation}
The general solution of this equation can be written by the superposition of discrete unbound states and continuous unbound states as
\begin{equation}
\psi=\sum_{nlm}c_{nlm}\psi_{nlm}+\int dk\sum_{lm}c_{lm}(k)\psi_{klm}~.
\end{equation}

The bound state solution can be written as 
\begin{equation}
\psi_{nlm}=e^{-i(\omega_{nlm}-\mu) t}R_{nl}(r)Y_{lm}(\theta,\varphi)~.
\end{equation}
In particular, the radial wavefunction that is regular at the origin is
\begin{equation}
R_{nl}(r)=\sqrt{\left(\frac{2}{n r_{0}}\right)^3\frac{(n-l-1)!}{2n(n+l)!}}e^{-\tilde{r}/n}\left(\frac{2\tilde{r}}{n}\right)^lL_{n-l-1}^{2l+1}\left(\frac{2\tilde{r}}{n}\right)~,
\end{equation}
where 
\begin{equation}
\tilde{r}=\frac{r}{r_{0}}~, \quad r_{0}=\frac{1}{\mu\alpha}~.
\end{equation}
$L_{p}^{q}(x)$ is the generalized Laguerre polynomial, defined by
\begin{equation}
L_{p}^{q}(x)=\sum_{n=0}^{p}(-1)^{n}\left(\begin{array}{c} p+q \\ p-n \end{array}\right)\frac{x^n}{n!}~.
\end{equation}

The unbound state solution can be written by
\begin{equation}
\psi_{klm}=e^{-i(\omega(k)-\mu) t}R_{kl}(r)Y_{lm}(\theta,\varphi)~.
\end{equation}
The radial wavefunctions which are regular at the origin and are outgoing at the infinity are, respectively, 
\begin{align}
&R^{0}_{kl}(r)=\frac{F_{l}(-1/kr_{0},kr)}{r}\to\left\{\begin{array}{ll}
C_{l}(-1/kr_{0})k(kr)^{l} & (r\to 0) \\
\frac{\sin{kr}}{r} & (r\to\infty)
\end{array}\right.~, \\
&R^{+}_{kl}(r)=\frac{H_{l}^{+}(-1/ka_{0},kr)}{r}\to\frac{e^{ikr}}{r}\ \ (r\to \infty)~,
\end{align}
where
\begin{gather}
F_{l}(\eta,\rho)=C_{l}(\eta)\rho^{l+1}e^{-i\rho} {}_{1}F_{1}(l+1-i\eta,2l+2,2i\rho)~, \\
C_{l}(\eta)=\frac{2^{l}e^{-\pi\eta/2}|\Gamma(l+1+i\eta)|}{(2l+1)!}~.
\end{gather}
$ {}_{1}F_{1}$ is the confluent hypergeometric function, and see~\cite{mathematicalfunction} for $H_{l}^{+}$, 
whose explicit form is not necessary in this paper.

\section{Frequency shift due to the angular momentum of the axion cloud}\label{AppB}
In this appendix, we derive the frequency shift due to the angular momentum of the axion cloud itself. This shift is sourced by the $t$-$\varphi$ component of the metric perturbation from the background as~\cite{Baumann:2018vus}
\begin{equation}
\left(i\partial_t+\frac{1}{2\mu}\nabla^2+\frac{\alpha}{r}\right)\psi=i\delta g^{t\varphi}\partial_{\varphi}\psi~.
\end{equation}
Writing the frequency $\omega=\bar{\omega}+\delta\omega$, we have
\begin{equation}
\delta\omega=\frac{-m\int d^{3}x ~\delta g^{t\varphi}|\psi_{nlm}(x)|^2}{\int d^{3}x|\psi_{nlm}(x)|^2}~.
\end{equation}
If we consider the leading term of the Kerr metric $\delta g^{t\varphi}=-2Ma/r^{3}$, we can reproduce the fifth tern of Eq.(\ref{Ene}). Here, we evaluate the metric perturbation due to the self-gravity of the axion cloud.  Approximating the background by Minkowski spacetime, we calculate the stationary $t$-$\varphi$ component of the metric perturbation by
\begin{equation}\label{lein}
\Delta \delta g_{ti}\simeq-16\pi T_{ti}~,
\end{equation}
where $\Delta$ is the Laplacian in flat space and $T_{ti}$ is the $t$-$i$ component of the energy momentum tensor. Here, we write $x^{i}=(x,y,z)$.
Solving Eq.(\ref{lein}), we have
\begin{equation}
\delta g_{t\varphi}=-16\pi\frac{\partial x^{i}}{\partial \varphi}\int d^{3}x'\frac{-1}{4\pi|\bm{x}-\bm{x}'|}T_{ti}~.
\end{equation}
On the other hand, the $t$-$\varphi$ component of the energy momentum tensor is
\begin{equation}
T_{t\varphi}=-\frac{2\omega m}{\mu}|\psi_{nlm}(x)|^2~
\end{equation}
and we can evaluate the angular momentum of the axion cloud as
\begin{equation}
J_{\mathrm{c}}\simeq -\int d^{3}x T_{t\varphi}=\frac{2\omega m}{\mu}\int d^{3}x|\psi_{nlm}(x)|^2~.
\end{equation}
As a result, using $\delta g^{t\varphi}=\delta g_{t\varphi}/r^2\sin^2\theta$, we have the frequency shift as
\begin{align}
\frac{\delta\omega}{\mu}&=\frac{4m}{\mu}J_{\mathrm{c}}\frac{\int d^{3}xd^{3}x'\frac{1}{r\sin\theta}\frac{1}{r'\sin\theta'}\frac{\cos(\varphi-\varphi')}{|\bm{x}-\bm{x}'|}|\psi_{nlm}(x)|^2|\psi_{nlm}(x')|^2}{\left(\int d^{3}x|\psi_{nlm}(x)|^2\right)^2} \notag \\
&=4m\alpha^5\frac{J_{\mathrm{c}}}{M^2}\sum_{LM}\frac{4\pi}{2L+1} \notag \\
& \quad \times\int d\tilde{r}d\tilde{r}'r\tilde{r}\tilde{R}_{nl}^2(\tilde{r})\tilde{R}_{nl}^2(\tilde{r}')\left(\frac{\tilde{r}'^{L}}{\tilde{r}^{L+1}}\theta(\tilde{r}-\tilde{r}')+\frac{\tilde{r}^{L}}{\tilde{r}'^{L+1}}\theta(\tilde{r}'-\tilde{r})\right) \notag \\
& \quad \times\int d\theta d\varphi d\theta' d\varphi'\cos(\varphi-\varphi')Y_{LM}(\Omega)|Y_{lm}(\Omega)|^2Y^{\ast}_{LM}(\Omega')|Y_{lm}(\Omega')|^2~.
\end{align}
In particular, for the cloud with $\ket{nlm}=\ket{21\pm1}$, we have
\begin{equation}
\frac{\delta\omega}{\mu}=\frac{7}{384}m\alpha^5\frac{J_{\mathrm{c}}}{M^2}~.
\end{equation}
In this case, the parameter we defined in Sec.~\ref{BRh} become $x=(7/384)/(1/12)=7/32$.
This parameter $x$ depends on the choice of the pair of levels, the one that produces the metric perturbation and the one whose energy eigenvalue is affected by it. However, since $x$ remains to be $\mathcal{O}(1)$ anyway, 
we adopt this value as an approximation for our analysis. 

Note that the frequency shift due to $t$-$t$ component of the self-gravity is also calculated in the appendix of Ref.~\cite{Baryakhtar:2020gao}. 
This effect comes from the order of $\mu\alpha^3$. Hence, when we consider the transition between different $n$ modes, which is the order of $\mu\alpha^2$, we can neglect it.

\bibliographystyle{ptephy}
\bibliography{ref}

\end{document}